\documentclass[a4paper, 11 pt]{article}
\usepackage[T1]{fontenc}
\usepackage[utf8]{inputenc}
\usepackage[english]{babel}
\usepackage{amsmath}
\usepackage{amssymb}
\usepackage{braket}
\usepackage{mathtools}
\usepackage{dsfont}
\usepackage{subcaption}
\usepackage{booktabs}
\usepackage{yfonts}
\usepackage{setspace}
\usepackage{cite}
\usepackage{verbatim}

\usepackage{amstext} 
\usepackage{array}   
\newcolumntype{C}{>{$}c<{$}} 

\usepackage{geometry}
\usepackage[usenames,dvipsnames]{color}
\usepackage{hyperref}
\hypersetup{
  colorlinks,
  citecolor=Blue,
  linkcolor=Blue,
  urlcolor=Blue}

\numberwithin{equation}{section}

\def\be{\begin{equation}}
\def\ee{\end{equation}}

\newcommand{\diff}{\mathrm{d}}
\newcommand{\ii}{\mathrm{i}} 

\def\DD{\alpha}
\def\II{I}
\def\CC{{\mathcal{C}}}

\begin{document}

\pagestyle{empty}

\begin{center}

$\,$
\vskip 2cm

{\LARGE{\bf The black hole at the end of the cone:}}\\[3mm] {\Large{\bf localizing the anomaly polynomial on toric geometries}}

\vskip 1cm

{\large
Davide Cassani,${}^{\textrm a}$
Enrico Turetta${}^{\textrm b}$
}

\vskip 1cm

\end{center}

\renewcommand{\thefootnote}{\arabic{footnote}}

\begin{center}
$^{\textrm a}$ {\it INFN, Sezione di Padova, Via Marzolo 8, 35131 Padova, Italy},\\ [2mm] 
$^{\textrm b}${\it Department of Physics and Research Institute of Basic Science, Kyung Hee University, Seoul 02447, Republic of Korea}.

\vskip 3cm

 {\bf Abstract} 
\end{center}

{\noindent 
We consider five-dimensional supergravity coupled to vector multiplets, gauged or ungauged, and propose an efficient method to evaluate the on-shell action of supersymmetric black saddle solutions with toric ${\rm U}(1)^3$ symmetry and general topology, explicitly known or just assumed to exist, including higher-derivative corrections. This is equivalent to equivariant integration of the anomaly polynomial six-form over simplicial cones obtained by decomposing the toric diagram of the solution. The on-shell action is then expressed as a sum over contributions localized at the tip of each cone. We obtain a simple derivation of recently calculated expressions as well as new predictions, both for the on-shell action of the black saddles and the Wald entropy of the related extremal solutions. These may be asymptotically AdS$_5$ or asymptotically flat. As examples, we discuss black holes, black rings and black lenses, including a black hole in the background of a topological soliton. For a putative supersymmetric AdS$_5$ black ring, we match our entropy prediction with the Bekenstein-Hawking entropy computed from the known near-horizon solution.
}


\newpage
\setcounter{page}{1}
\pagestyle{plain}

\tableofcontents

\newpage 

\section{Introduction}

A central open problem in quantum gravity is to understand the gravitational path integral as a sum over geometries. This requires investigating  questions such as which spacetime topologies should contribute, to what extent complex saddles are admissible, how should we interpret wormhole configurations, and how does the gravitational path integral emerge from a fundamental microscopic description.

Given the difficulty of
 these questions, it is natural to consider settings in which the path integral is sufficiently constrained to become tractable. A particularly convenient framework is provided by supergravity with supersymmetric boundary conditions. If Euclidean time is compactified on a circle, the path integral admits a trace interpretation and, in the presence of supersymmetry, computes a protected quantity known as the gravitational index, see e.g.~\cite{Cassani:2025sim} for a brief review. Furthermore, by imposing asymptotically AdS boundary conditions, one can exploit holography and dual SCFT indices to gain further insight and make detailed comparisons, see e.g.~\cite{Kinney:2005ej,Cabo-Bizet:2018ehj,Choi:2018hmj,Benini:2018ywd,Benini:2015eyy,Zaffaroni:2019dhb} for some representative references.

The saddles contributing to the gravitational index include complexified black hole solutions that are supersymmetric but non-extremal~\cite{Cabo-Bizet:2018ehj}. More generally, one may expect contributions from other topologies. Determining the role of these other saddles is an important step toward understanding the microscopic origin of the index and, more broadly, the structure of the gravitational path integral.

Five-dimensional supergravity provides a rich setting in which to investigate these questions. In asymptotically flat Lorentzian spacetime, a diverse landscape of supersymmetric extremal solutions is known explicitly, including black holes, black rings, black lenses, topological solitons, and composite configurations thereof, see the classification in~\cite{Breunholder:2017ubu}. By contrast, the construction of supersymmetric solutions in asymptotically AdS$_5$ backgrounds has proven considerably more challenging. At present, the catalogue of solutions is essentially limited to black holes with spherical horizons \cite{Gutowski:2004ez,Gutowski:2004yv,Chong:2005hr,Kunduri:2006ek} and a very small number of topological solitons~\cite{Chong:2005hr,Cassani:2015upa,Durgut:2021bpk}. Understanding whether additional supersymmetric geometries exist, and how they contribute to the gravitational index, therefore remains an open problem.

Even when a supersymmetric solution with a given topology is not known explicitly, but is only assumed to exist, it is still valuable to determine its on-shell action and hence its possible contribution to the gravitational path integral. In particular, comparing its action with those of other saddles allows one to assess whether it can compete with, or even dominate, the partition function and thereby give rise to a new phase.

Remarkably, recent developments in supergravity, starting with~\cite{BenettiGenolini:2023kxp,BenettiGenolini:2023ndb}, have shown that precisely this information can be extracted using the tool of equivariant localization. Without requiring the explicit construction of a solution, localization determines the supersymmetric on-shell action from data associated with the fixed loci of the relevant isometries.  This approach is especially powerful when one aims at incorporating higher-derivative corrections, where exact solutions are even harder to obtain \cite{Hristov:2024cgj,BenettiGenolini:2026qdm,Gaar:2026nqq}.

In this paper, we start from five-dimensional Fayet-Iliopoulos gauged supergravity, namely minimally supersymmetric five-dimensional supergravity coupled to $n_{\rm v}$ vector multiplets, and with a gauging of the R-symmetry. This theory has a supersymmetric AdS$_5$ vacuum. We consider supersymmetric  non-extremal asymptotically AdS$_5$  solutions, with the Euclidean time compactified to a circle of finite length. We also assume the solutions have ${\rm U}(1)^3$ symmetry, given by combining evolution along the Euclidean time with rotation in the asymptotic spatial $S^3$. These are generically complex field configurations, of the type discussed in~\cite{Cabo-Bizet:2018ehj}. They are sometimes called black saddles, since they represent saddle-points of the gravitational path integral that computes a supersymmetric index, describing the contribution of black holes and other objects with an event horizon. We will also consider the limit where the gauging is turned off, which leads to ungauged supergravity with a Minkowski vacuum and asymptotically flat solutions.

We propose that the background-subtracted on-shell action of these solutions is efficiently calculated by formally extending the five-dimensional action to a six-dimensional space $\mathcal{M}_6$ with an underlying toric geometry,  decomposing this toric geometry into simplicial cones, and using equivariant localization to reduce the integral to a sum over isolated fixed point contributions, where the fixed points are just the tips of the cones. Noting that this is formally equivalent to integrating the anomaly polynomial associated with the theory allows us to incorporate higher-derivative corrections into the formula for the on-shell action.
 
In the following we illustrate our approach in some more detail. We proceed in three steps, that we denote ascent, refinement, and localization.

\paragraph{Ascent.} 
Consider the supergravity Lagrangian including higher-derivative terms. This can be systematically constructed using the off-shell conformal supergravity formalism \cite{Hanaki:2006pj, 
Ozkan:2013uk, Ozkan:2013nwa, Bobev:2021qxx,Gold:2023ymc,Hristov:2025ygn}. 
An effective (bosonic) Lagrangian valid at first order in the higher-derivative corrections has been given in~\cite{Cassani:2024tvk}. 
Here we will need just a minimal piece of information about the Lagrangian, namely that it contains the Chern-Simons terms
\be\label{CSterms_intro}
\mathcal{L} \, =\, - \frac{1}{24\pi^2}\left( k_{IJK} A^I \wedge \diff A ^J \wedge \diff A^K - \frac{1}{8}  k_I A^I \wedge R_{\alpha\beta} \wedge R^{\alpha\beta} \right)  + \ldots\,,
\ee
where $A^I$, $I=1,\ldots,n_{\rm v}+1$ are Abelian gauge fields and $R_{\alpha\beta}$ is the Riemann curvature two-form.
Moreover,  $k_{IJK}$ is a fully symmetric constant tensor that at the two-derivative level completely determines the  supergravity Lagrangian, while $k_I$ is a coupling controlling the four-derivative terms.\footnote{We define the Lagrangian $\mathcal{L}$ such that the Lorentzian action is $I_{\rm Lor} = \int \mathcal{L}$, hence we are absorbing the usual factor of $\frac{1}{16\pi G}$  in the couplings $k_{IJK}$ and $k_I$. 
} 
These terms are also those that reproduce the ${\rm U}(1)_I$ cubic and linear 't Hooft anomalies of dual $\mathcal{N}=1$ SCFT's.

Supersymmetric solutions have a Killing vector $\xi$, defined as a bilinear of the Killing spinor. We will focus on the class of solutions where $\xi$ is timelike in Lorentzian signature. Starting from this setup, we allow for analytic continuation to a complexified Euclidean setup. Recently, it has been shown in~\cite{Colombo:2025ihp,Colombo:2025yqy} that the two-derivative supergravity action evaluated on asymptotically AdS$_5$ supersymmetric solutions can be written as a Chern-Simons term,
 \be\label{5d_CS}
I_{2\partial}\,=\, -\frac{\ii}{24\pi^2} \int_{\mathcal{M}_5} k_{IJK} \,\eta^I\wedge \diff\eta^J \wedge \diff \eta^K \,.
\ee
 The local one-form $\eta^I$ is a shifted gauge potential $\eta^I = -A^I +\ldots$, determined by the solution and such that its exterior derivative is basic with respect to the foliation defined by the supersymmetric Killing vector, $\iota_\xi \diff\eta = 0$. The factor of $\ii$ arises from the continuation to Euclidean signature.

The on-shell action above is background-subtracted. Namely, assuming the solution is asymptotically (globally) AdS$_5$, the (regularized) on-shell action is renormalized by subtracting the (regularized) action of empty AdS$_5$ with the same coordinate identifications. This is interpreted as removing the contribution of the vacuum. 
There is an interesting story~\cite{Colombo:2025ihp,BenettiGenolini:2025icr,Park:2025fon,Colombo:2025yqy} about boundary terms whose outcome we are incorporating here and that can be summarized as follows.
Rewriting the bulk supergravity action as a Chern-Simons term also produces a set of locally exact terms which, under suitable regularity assumptions, take the form of boundary terms. Furthermore, application of the localization theorem (which will be one of our next steps too) yields additional boundary terms \cite{Couzens:2024vbn,Cassani:2024kjn}. 
 Remarkably, it turns out that after background subtraction, the full set of boundary terms  combines to give a vanishing contribution once the cutoff is removed~\cite{Colombo:2025ihp,BenettiGenolini:2025icr,Park:2025fon,Colombo:2025yqy}.
  In a sense that it would be interesting to make precise, background subtraction in this context can be understood as gluing (with opposite orientation) the solution of interest with empty AdS$_5$ along their (regularized) asymptotic boundaries, in such a way that the spacetime is effectively compact.
In the following we assume that all these steps have been taken and we will aim at evaluating the bulk action \eqref{5d_CS}, where $\mathcal{M}_5$ denotes the effectively compact background-subtracted geometry.
 
Evaluating Chern-Simons terms has the usual subtleties related with non-invariance under large gauge transformations and the need to work patchwise when the gauge connection is not globally well-defined.
The unambiguous way to define these terms is to uplift to a space with one more dimension. Namely, we consider a six-dimensional space $\mathcal{M}_6$ such that $\mathcal{M}_5=\partial\mathcal{M}_6$ and an extension of the five-dimensional solution to $\mathcal{M}_6$, so that the action can be written as
 \be
I_{2\partial}\,=\, -\frac{\ii}{24\pi^2} \int_{\mathcal{M}_6} k_{IJK} \, \diff\eta^I\wedge \diff\eta^J \wedge \diff \eta^K\,,
\ee
where now $\eta^I$ is the extension of the connection one-form to $\mathcal{M}_6$.

Although the Chern-Simons reformulation of the supersymmetric on-shell action has been proven at the two derivative level only, we conjecture with no proof that the same structure extends to the higher-derivative level, namely that
  the higher-derivative supersymmetric on-shell action is captured by
 \be
 \label{eq:CS_HD}
I\,=\,  -\frac{\ii}{24\pi^2}\int_{\mathcal{M}_6} \left( k_{IJK}\, \diff\eta^I\wedge \diff\eta^J \wedge \diff \eta^K - \frac{1}{8}\,k_I \diff \eta^I\wedge R_{\alpha\beta} \wedge R^{\alpha\beta}\right)\,.
 \ee

We note that the expression above is closely related to the anomaly polynomial associated with the supergravity Chern-Simons terms, with the difference that  $\diff\eta^I$ cohomology classes are used instead of those of the ${\rm U}(1)_I$ gauge bundle alone. This is also the 't Hooft anomaly polynomial for the four-dimensional SCFT living at the AdS$_5$ boundary.\footnote{ In the ungauged case there is no dual SCFT but anomalies are nevertheless known to play a central role in evaluating supersymmetric on-shell actions, see e.g.~\cite{Kraus:2005vz}. 
 }

\paragraph{Refinement.} Assume a supersymmetric non-extremal solution on $\mathcal{M}_5$, with ${\rm U}(1)^3$ symmetry and a given topology.
In order to evaluate the action, we should identify a suitable $\mathcal{M}_6$ extending $\mathcal{M}_5$. We exploit the fact that solutions with ${\rm U}(1)^3$ isometry can be treated borrowing techniques from toric geometry, as explained in~\cite{Colombo:2025ihp,Colombo:2025yqy,Martelli:2023oqk} (see also~\cite{Cassani:2025iix} for the asymptotically flat case).
Essentially, we can think of $\mathcal{M}_6$ as the extension naturally defined by the toric data of $\mathcal{M}_5$ and carrying the same toric structure together with the action of $\xi$, as follows.
Modding out by the torus action, one can represent the geometry as a three-dimensional polytope with facets, edges and vertices given by loci where the action of one, two or three independent ${\rm U}(1)$'s vanish, respectively. 
To each ${\rm U}(1)$ that degenerates we can associate a vector with three integer components, normal to the facet in the polytope, specifying the linear combination of the three ${\rm U}(1)$'s in the torus that collapses. The collection of all such vectors forms a fan. We choose the fan that is defined by $\mathcal{M}_5$. When more than three facets meet at a vertex, the space is singular and needs a regularization. Indeed, in order to apply the localization theorem we need $\mathcal{M}_6$ to be smooth up to orbifold singularities. In order to obtain this, we look for a refinement of the fan, which amounts to decomposing the fan into triplets of linearly independent vectors, so that each triplet generates a simplicial cone. A simplicial cone is a cone such that the tip is either smooth, or an orbifold singularity. This gives our resolution.

The procedure is standard in toric geometry but our construction will not be equally rigorous, since not all of the usual conditions for toric geometry are going to be satisfied in the solutions of interest. Also, in the examples we will study  we will encounter rather unfamiliar geometries, such as degenerate fans and non-convex polytopes.

We will also need global information about the gauge fields at the fixed loci of the ${\rm U}(1)^3$ action on $\mathcal{M}_5$, which are inherited by $\mathcal{M}_6$.

\paragraph{Localization.} Having specified the necessary geometric data of $\mathcal{M}_6$, we can now calculate the action integral using the Berline-Vergne-Atiyah-Bott (BVAB) equivariant localization theorem~\cite{BerlineVergne,Atiyah:1984px}. In order to do so, we equivariantize the integrand by constructing an equivariantly closed polyform on $\mathcal{M}_6$. We localize with respect to the action of the  supersymmetric Killing vector $\xi$. Then the integral localizes to a sum over contributions from the fixed points of $\xi$. These are just the endpoints of the different cones in the decomposition of the fan. 
 Formally, this computation is the same as equivariant localization of the anomaly polynomial over the full six-dimensional ambient space, that we present separately since it may be of independent interest.

The final formula for the on-shell action is a sum over the contribution from the tip of each of the cones in the decomposition. The contribution at each fixed point depends on the weights of the action of $\xi$ with respect to the  triplet of fan vectors that determines the cone, together with the global gauge data. Ultimately, the action is expressed in terms of a few data: the angular velocities $\omega_1,\omega_2$ and the electrostatic potentials $\varphi^I$ specified at the asymptotic boundary, which are the continuous variables entering in the action seen as a grand-canonical thermodynamic potential, together with a set of integer parameters that characterize the topology of the solutions under study.
The general formula is given in eqs.~\eqref{eq:result_action}--\eqref{eq:result_action_4} below. 

 It may be useful to comment on the relation between our approach and other works using equivariant localization to evaluate the five-dimensional supergravity on-shell action. In~\cite{Cassani:2024kjn} equivariant localization is introduced in five-dimensional (ungauged) supergravity, though supersymmetry is perhaps not exploited in the most efficient way. In~\cite{Colombo:2025ihp,Colombo:2025yqy}, the five-dimensional action is reduced along the supersymmetric Killing vector $\xi$ and localization is then implemented using a different Killing vector. 
 A general formula for the two-derivative on-shell action is obtained, expressed in terms of a sum over all fan vectors.
Our results agree with this formula in all examples where we could make a comparison. It would be interesting to establish a more general proof. Moreover, our expression appears slightly simpler as we repackage the fan vector contributions in cone contributions, where each cone corresponds to a triplet of vectors.  In addition, we incorporate higher derivatives.
Yet a different approach, used in \cite{BenettiGenolini:2025icr,Park:2025fon,Gaar:2026nqq} in order to recover the spherical black hole on-shell action, is to dimensionally reduce to four dimensions along a non-susy ${\rm U}(1)$ Killing vector and then localize there with respect to $\xi$.  Instead of reducing to four dimensions and using an auxiliary Killing vector in addition to the supersymmetric Killing vector, in this paper we have taken the ascending path to six dimensions before localizing down to zero dimensions by just exploiting the supersymmetric Killing vector.

\paragraph{Examples.} 

In order to corroborate and illustrate our method, we discuss a number of relevant examples and compare the results with the existing literature, with a focus on higher-derivative corrections when they are known. In all cases, we reproduce the two derivative results in~\cite{Colombo:2025yqy} and extend them with the higher-derivative contributions. 

We present the on-shell action for black saddles with spherical ($S^3$), lens space ($S^3/\mathbb{Z}_n$) and ring ($S^2\times S^1$) horizon topology, including a discrete family from shifts and orbifolds, and a black hole in the background of a topological soliton.

Given the on-shell action, we can Legendre transform it to find the Wald entropy of the corresponding supersymmetric extremal solution, which has a Lorentzian interpretation.
We compare with explicitly known asymptotically flat solutions and, for the lens and ring horizon topologies, comment on their possible realization in AdS, which is presently unknown. We highlight the cases where our prediction for the higher-derivative corrections is new. We also compare our expression for the entropy of a putative AdS$_5$ black ring with the Bekenstein-Hawking entropy of the explicit near-horizon solution constructed in~\cite{Kunduri:2007qy}, finding a precise agreement.

Our findings therefore establish a unified approach to the evaluation of supergravity on-shell actions in both gauged and ungauged settings, including higher-derivative corrections. This perspective emphasizes the underlying unity among the various solutions that contribute as saddles to gravitational indices.

\paragraph{Plan of the paper.} The remainder of the paper is organized as follows. In section~\ref{sec:equivariant_details}, we present the equivariant integration of the anomaly polynomial over the toric geometries of interest and show how the result localizes to fixed point contributions. In section~\ref{sec:sugra_sol}, we review the relevant properties of the supersymmetric solutions, illustrating the global data required for our analysis. Section~\ref{sec:OSAfromAP} contains our main results: we give our formula for the on-shell action, including higher-derivative corrections, and apply it to examples. In section~\ref{sec:compare_near_horizon}, we match our prediction for the AdS$_5$ black ring entropy with the entropy extracted from the known near-horizon geometry. Finally, in section~\ref{sec:conclusions}, we comment on our findings and outline directions for future research. Appendix~\ref{appentropy} contains details of the Legendre transform of the on-shell action that yields the Wald entropy.


\section{Equivariant integration of anomaly polynomial}
\label{sec:equivariant_details}

We start by presenting our main tool, namely  equivariant integration of the anomaly polynomial six-form that is associated with the supergravity Chern-Simons terms.

Equivariant integration of the anomaly polynomial on some submanifold (such as a Riemann surface) is a well-established technique in field theory and holography, yielding the anomaly polynomial associated with the lower-dimensional field theory that arises by compactifying the parent theory on that submanifold, see e.g.~\cite{Benini:2009mz,Bah:2012dg,Bah:2019rgq,Hosseini:2020vgl}. 
In the following, we will be interested in integrating the anomaly polynomial over the whole ambient space, namely we will localize the integral to zero dimensions. This type of computation has also already appeared in the literature, in particular as a tool to evaluate supersymmetric Casimir energies~\cite{Bobev:2015kza} and other saddle-point contributions to supersymmetric field theory partition functions in Cardy-like limits~\cite{Ohmori:2021dzb,Cassani:2024tvk}.\footnote{The results of~\cite{Cassani:2024tvk} already show that the equivariant integration of the anomaly polynomial, originally carried out to evaluate a Cardy-like limit of the SCFT index, also computes the on-shell action of the dual black hole saddle in supergravity. In the present paper, we further notice that this follows from the general form~\eqref{eq:CS_HD} of the on-shell action.}
 However, the integration manifold considered in these applications is the simplest toric geometry, namely flat space. In the following, we extend the method to general toric geometries with non-trivial topology, focusing on six-dimensional spaces. See also~\cite{Hristov:2026tde} for a related recent application in two dimensions more. 
 
The relevant anomaly polynomial $\mathtt P$ is a six-form defined on ${\cal M}_6$, given by
\begin{equation}
\label{eq:P}
\mathtt P \,=\, \frac{1}{6}k_{IJK} \mathtt c_1\left(\frac{F^I}{2\pi}\right)\mathtt c_1\left(\frac{F^J}{2\pi}\right) \mathtt c_1\left(\frac{F^K}{2\pi}\right) - \frac{1}{24}k_I \mathtt c_1\left(\frac{F^I}{2\pi}\right)\mathtt p_1\left(T{\cal M}_6\right)\,,
\end{equation}
where $k_{IJK}$ and $k_I$ are respectively cubic and linear 't Hooft anomaly coefficients, while $\mathtt c_1\left(F^I/2\pi\right)$ denotes the first Chern class associated to the $U(1)_I$ gauge symmetry.\footnote{In our supergravity application, we will replace $F^I$ with $-\diff\eta^I$.} Finally, $\mathtt p_1$ is the first Pontryagin class of $T{\cal M}_6$, which in our conventions is expressed in terms of the curvature two-form $ R_{\alpha\beta}$ as 
\begin{equation}
\mathtt p_1 = \frac{1}{8\pi^2} R_{\alpha\beta} \wedge  R^{\alpha\beta}\,.
\end{equation}

These cohomology classes are the extension to ${\cal M}_6$ of cohomology classes originally defined in one (or two, for field theory applications) dimensions less.
We now need to specify the backgrounds ${\cal M}_6$ relevant for our applications.


\subsection{Elements of toric geometry of complex dimension three}
\label{sec:toric_geom}

In this section we review a few properties of six-dimensional toric orbifold geometries that will be needed in the following, mostly following~\cite{Martelli:2023oqk}. Further details can be found e.g.\ in the lectures~\cite{Bouchard:2007ik,Closset:2009sv} and in~\cite{Martelli:2005tp,Martelli:2006yb}. 

The geometry ${\cal M}_6$ is a three-torus fibration over the three-dimensional polytope 
\begin{equation}
{\cal P} = \{ y \in \mathbb R^3 \,:\,l_a = y\cdot V^a - \lambda_a \geq 0\}\,,\qquad a = 0,1,...,s\,,
\end{equation} 
whose shape is controlled by the parameters $\lambda_a$. The facets of ${\cal P}$ are defined by the equations $l_a =0$. To each facet one associates an inward-pointing normal vector $V^a$ in the lattice $\mathbb{Z}^3$, which identifies the U(1) subgroup of U(1)$^3$ whose action degenerates over the facet. The torus fibration is non-degenerate in the interior of ${\cal P}$, while the circle generated by $V^a$ collapses over the facet $l_a=0$. The inverse image in ${\cal M}_6$ of the facet is a complex codimension one subspace, known as divisor and denoted by ${\cal D}_a$. This means that the divisor ${\cal D}_a$ is the fixed locus of the U(1) action generated by the vector $V^a$. At the intersection of $m$ facets, more independent cycles of U(1)$^3$ degenerate, giving a subspace of complex codimension $m$. In particular, the intersection of three facets is a vertex of the polytope, corresponding to an isolated fixed point for the U(1)$^3$ action. The collection of vectors $\{V^a\}_{a=0,1...,s}$, together with the intersection data for the facets, specifies the fan of the space.

Each facet has an associated normal line bundle ${\cal L}_a$. A representative for the first Chern class of ${\cal L}_a$ is
\begin{equation}
\mathtt c_1 \left( {\cal L}_a\right) = \diff \left( \mu_a\cdot \diff \phi\right)\,,
\end{equation} 
where $\phi_i = \left(\phi_0,\phi_1,\phi_2\right)$ are $2\pi$-periodic angular coordinates on the torus and $\mu_a^i$ are suitable momentum maps defined by
\begin{equation}\label{eq:momentummap}
\iota_{\partial_{\phi_i}} \mathtt c_1\left({\cal L}_a\right) = - \diff \mu_a^i\,.
\end{equation} 
The momentum maps are normalized as
\begin{equation}
\sum_{a=0}^s\mu^i_aV^a_j = - \frac{\delta^i_j}{2\pi}\,.
\end{equation}
The line bundles are not all independent, since combining the previous equations gives a linear relation: 
\begin{equation}
\label{eq:line_bundle}
\sum_{a=0}^sV^a_i\mathtt c_1 \left({\cal L}_a\right) =0\,.
\end{equation}
A representative for the first Chern class associated with a ${\rm U}(1)_I$ symmetry can be expanded in this basis as\footnote{The sign is chosen so that, upon the replacement $F^I \to -\diff \eta^I$ of section~\ref{general_formula_action}, the $\alpha^I_a$ coincide with the coefficients in the expansion~\eqref{eq:deta} of $\diff \eta^I$.}
\begin{equation}
\label{eq:gauge_chern}
\mathtt c_1 \left(\frac{F^I}{2\pi}\right)\, =\, -\sum_{a=0}^s \DD_a^I\,\mathtt c_1\left( {\cal L}_a\right)\,.
\end{equation}

In this context, of special interest are toric cones, in particular Calabi-Yau cones. The Calabi-Yau condition requires all vectors $V^a$ to lie in a common plane. Then, after a suitable ${\rm SL}(3,\mathbb Z)$ rotation, one obtains that the first components of all vectors $V^a$ are $1$, 
\begin{equation}
V^a = \left(1, v^a\right) \,,\qquad v^a \in \mathbb Z^2\,.
\end{equation}

Toric geometries often have singularities worse than orbifold singularities. In particular, this occurs when more than three facets meet at a vertex. 
To obtain a space with at worst orbifold singularities, one can consider a toric resolution, obtained by replacing the singular fan by a  refinement into simplicial cones.
This means that the original fan is decomposed into three-dimensional cones, each cone being associated to a triplet of vectors $V^a$ in the fan.\footnote{If this condition cannot be met, we can always resolve the fan by adding one (or more) additional triangulation vectors. However, we will not deal with examples of this type here.} 
We will denote by 
\be
\CC_k= \{ V^{k_1},V^{k_2}, V^{k_3}\}
\ee
 the simplicial cones in the refinement, where $k = 1,...,n_\CC$, and $n_\CC$ is the number of cones. 
 The apex of each simplicial cone is a fixed point of the torus action and the local geometry around it is a finite abelian quotient of $\mathbb C^3$.  In particular, whenever the Calabi-Yau condition is satisfied the refinement amounts to a triangulation of the two-dimensional fan $\{v^a\}_{a=1,...,s}$, where each triangle corresponds to a simplicial cone.


\paragraph{Example: small resolution of the conifold singularity.}

\begin{figure}[!htb]
	\centering
	\includegraphics[width=0.3\textwidth]{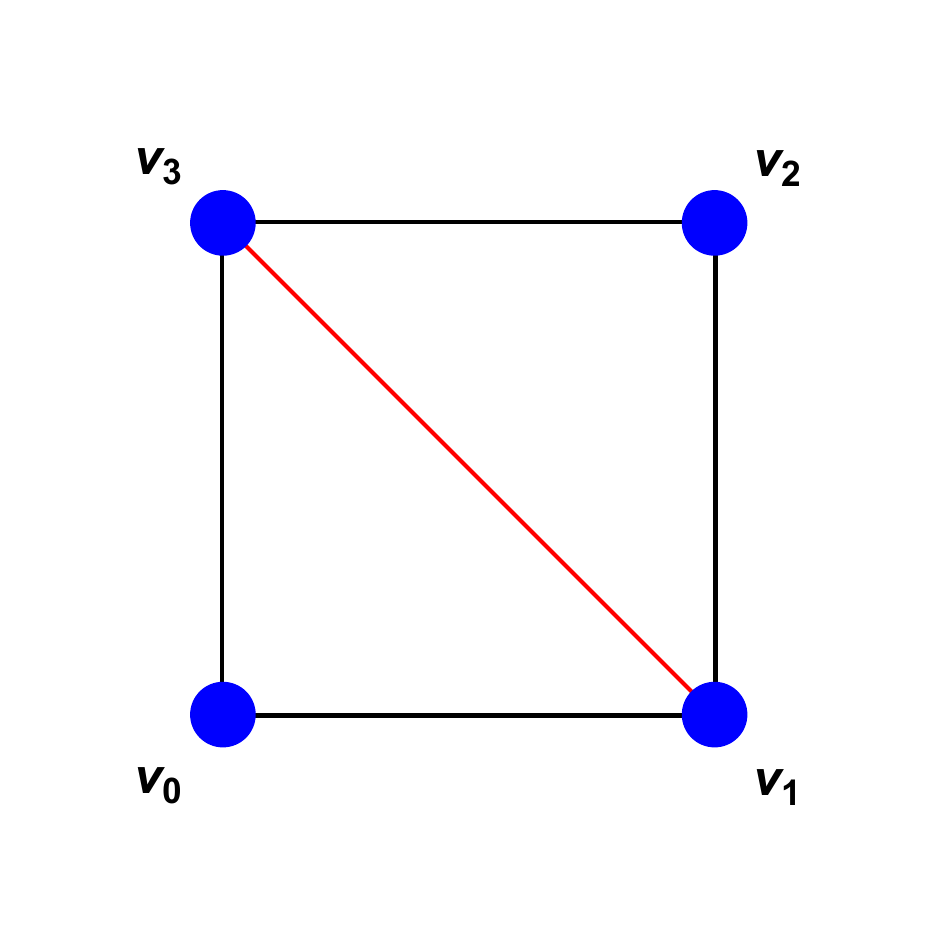}
	\caption{\it  Toric diagram	of the conifold. The diagonal specifies one of the two possible small resolutions.}
	\label{fig:diagram_conifold}
\end{figure}

We can use the simple well-known example of the conifold to illustrate the procedure. The toric data of the conifold are given by the vectors
\begin{equation}
V^0 = \left(1,0,0\right)\,,\qquad V^1 = \left(1,1,0\right)\,,\qquad V^2 = \left( 1,1,1\right)\,,\qquad V^3 = \left(1,0,1\right)\,.
\end{equation} 
This defines a toric Calabi-Yau three-fold, since the Calabi-Yau condition is satisfied. The corresponding cone is singular with four facets meeting at its tip. Therefore, we need to resolve the conifold singularity, by triangulating the fan. To do so, let us introduce the two-dimensional vectors
\begin{equation}
v^0 = \left(0,0\right)\,,\qquad v^1 = \left( 1,0\right)\,,\qquad v^2 = \left( 1,1\right)\,,\qquad v^3 = \left( 0,1\right)\,.
\end{equation}
Joining the vector endpoints gives the toric diagram, which in this case is a square,
see Fig.~\ref{fig:diagram_conifold}. 

Adding the diagonal connecting $v^1$ and $v^3$, the refined fan is obtained by replacing the original fan with the two cones
\begin{equation}
\label{eq:cones_conifold}
\CC_1 = \{V^0, V^1, V^3\} \quad \text{and} \quad \CC_2 = \{V^1, V^2, V^3\}\,.
\end{equation}
This is the small resolution of the conifold, in which the singular point is replaced by a compact holomorphic curve (see e.g.~\cite{Closset:2009sv} for further details). Considering the other diagonal leads to an equivalent resolution.


\subsection{Equivariant localization}
\label{sec:int_an_poly}

The BVAB localization theorem in equivariant cohomology allows one to express the integral of an equivariantly closed polyform in terms of contributions from the fixed points of the torus action. We work equivariantly with respect to the U(1)$^3$ action on ${\cal M}_6$, generated by the vectors $\partial_{\phi_i}$. Correspondingly, we introduce the equivariant parameters $\omega_i$ and the vector 
\be
\xi\, =\, \sum_{i=0}^2\omega_i\,\partial_{\phi_i}\,.
\ee 
The torus action on the resolved toric orbifold has $n_\CC$ isolated fixed points, corresponding to the tip of each simplicial cone. Consider a polyform $\hat\Psi$ on ${\cal M}_6$.\footnote{We denote by $\hat\Psi$ a polyform providing an equivariant extension of a form $\Psi$.} The latter is equivariantly closed if
\begin{equation}
\label{eq:equiv_diff}
\diff_\xi \hat\Psi \equiv \left(\diff  +\iota_\xi\right)\hat\Psi =0 \,.
\end{equation} 
The localization theorem then gives~\cite{Atiyah:1984px,BerlineVergne}
\begin{equation}
\label{eq:localization_formula}
\int_{\mathcal M_6} \hat\Psi \,\,=\,\, \sum_{k=1}^{n_\CC} \,\frac{\hat\Psi\big|_{y_k}}{d_k\,\hat e\big|_{y_k}}\,\,,
\end{equation}
where the sum runs over the fixed points $y_k$, while $d_k$ is the order of the orbifold singularity at the fixed point $y_k$ and $\hat e$ is the equivariant Euler class. 

In order to apply the theorem to the anomaly polynomial, we need to promote the characteristic classes entering in its definition \eqref{eq:P} to their equivariant extension, so that $\mathtt P$ becomes an equivariantly closed polyform. 
This is achieved by replacing
\begin{equation}
\label{eq:equivariant_chern}
\mathtt c_1 \left( {\cal L}_a\right)\  \to\ \hat {\mathtt c}_1 \left({\cal L}_a\right) = \mathtt c_1 \left( {\cal L}_a\right) + \omega \cdot \mu_a\,,
\end{equation}
where $\mu_a^i$ are the momentum maps \eqref{eq:momentummap}.
At the fixed point $y_k$ associated with the cone $\CC_k$,  assuming generic values of the equivariant parameters, the tangent space decomposes as a direct sum of three complex lines, denoted by ${\cal L}_{k_1}$, ${\cal L}_{k_2}$, ${\cal L}_{k_3}$. Therefore, the equivariant Euler class gives (see e.g.~\cite{BenettiGenolini:2023ndb})
\begin{equation}\label{eq:expression_euler}
\hat e\big|_{y_k} \,=\, \prod_{k_{1,2,3}}{\rm Pf}\left[ \frac{\hat{ \mathtt R}_{mn}}{2\pi}\right]\Big|_{{\cal L}_{k_i}} \,=\, \prod_{k_{1,2,3}}\hat {\mathtt c}_1\left({\cal L}_{k_i}\right)\Big|_{y_k} \,=\, \epsilon_{k_1}\epsilon_{k_2}\epsilon_{k_3}\,,
\end{equation}
where $\hat {\mathtt R}_{mn} = \mathtt R_{mn} + (L_\xi)_{mn}$ denotes the equivariant extension of the curvature two-form, and $\epsilon_{k_i} \equiv \omega\cdot \mu_{k_i}$ are the weights of the U(1) action generated by $\xi$ on the tangent space at $y_k$.\footnote{For non-generic values of the equivariant parameters, \eqref{eq:localization_formula} still applies but one should not use \eqref{eq:expression_euler}.} Equivalently, these weights are obtained by expanding $\xi$ in the basis of vectors degenerating at $y_k$,
\begin{equation}
\xi \,=\, \epsilon_{k_1}V^{k_1} + \epsilon_{k_2}V^{k_2} + \epsilon_{k_3} V^{k_3} \,,
\end{equation}
with
\begin{equation}
\label{eq:equiv_params}
\epsilon_{k_1}=\frac{\left[\xi,V^{k_2},V^{k_3}\right]}{\left[V^{k_1},V^{k_2},V^{k_3}\right]}\,,
\qquad
\epsilon_{k_2}=\frac{\left[V^{k_1},\xi,V^{k_3}\right]}{\left[V^{k_1},V^{k_2},V^{k_3}\right]}\,,
\qquad
\epsilon_{k_3}=\frac{\left[V^{k_1},V^{k_2},\xi\right]}{\left[V^{k_1},V^{k_2},V^{k_3}\right]}\,,
\end{equation}
where we introduced the notation
\begin{equation}
\left[V,W,Z\right]\,\equiv\, \det(V,W,Z)\,.
\end{equation}
The local orbifold order is instead given by\footnote{To be precise, the order of the local orbifold at \(y_k\) is $\big|\left[V^{k_1},V^{k_2},V^{k_3}\right]\big|$. However, in computing \eqref{eq:localization_formula}, we use the signed determinant to account for possible minus signs related to the orientation of the ordered basis $\{V^{k_1},V^{k_2},V^{k_3}\}$. 
}  
\begin{equation}
d_k \,=\,
\big[V^{k_1},V^{k_2},V^{k_3}\big]\,.
\end{equation}
The equivariant extension of the first Chern class \eqref{eq:gauge_chern}, obtained by the substitution \eqref{eq:equivariant_chern}, evaluated at the fixed point $y_k$ therefore takes the form
\begin{equation}
\label{eq:Delta}
\hat {\mathtt c}_1\left( \frac{F^I}{2\pi}\right)\Big|_{y_k} \,=\, -\sum_{i=1,2,3}\DD^I_{k_i}\epsilon_{k_i}\,.
\end{equation}
Similarly, the equivariant extension of the first Pontryagin class at the fixed point $y_k$ gives
\begin{equation}
\hat {\mathtt p}_1\left( T{\cal M}_6\right)\big|_{y_k} \,=\, \sum_{i=1,2,3} \left(\hat {\mathtt c}_1\left({\cal L}_{k_i}\right)\big|_{y_k}\right)^2 \,=\, (\epsilon_{k_1})^2 + (\epsilon_{k_2})^2 + (\epsilon_{k_3})^2\,.
\end{equation}

Using the ingredients above, the equivariant integral of the anomaly polynomial over the resolved toric geometry localizes onto a sum over isolated fixed point contributions, one for each cone in the resolution. We obtain: 
\begin{equation}
\label{eq:int_an_poly}
 2\pi \ii \int_{\mathcal M_6}\mathtt P \,=\, \sum_{\CC_k}\left(\frac{1}{6}\,k_{IJK} I^{IJK}(\CC_k) - \frac{1}{24}\,k_J I^J(\CC_k)\right)\,,
\end{equation}
where the contribution of a cone $\CC_k$ is
\begin{equation}\label{eq:int_an_poly_2}
I^{IJK}(\CC_k)\,=\, -2\pi \ii \, \frac{\big(\sum_{i=1}^3 \DD^I_{k_i}\epsilon_{k_i}\big)\,\big(\sum_{i=1}^3 \DD^J_{k_i}\epsilon_{k_i}\big)\,\big(\sum_{i=1}^3 \DD^K_{k_i}\epsilon_{k_i}\big)}{d_k\,\epsilon_{k_1}\epsilon_{k_2}\epsilon_{k_3}}\,,
\end{equation}
together with
\begin{equation}\label{eq:int_an_poly_4}
I^J(\CC_k)\,=\, -2\pi \ii\,\frac{\big(\sum_{i=1}^3 \DD^J_{k_i}\epsilon_{k_i}\big)\left[(\epsilon_{k_1})^2+(\epsilon_{k_2})^2+(\epsilon_{k_3})^2\right]}{d_k\,\epsilon_{k_1}\epsilon_{k_2}\epsilon_{k_3}}\,,
\end{equation}
and the equivariant parameters $\epsilon_{k_{1,2,3}}$ are defined by \eqref{eq:equiv_params}.


\section{Supergravity solutions with U(1)$^3$ symmetry}\label{sec:sugra_sol}

In this section we review some properties of asymptotically AdS$_5$ solutions of five-dimensional supergravity coupled to $n_{\rm v}$ vector multiplets, with a gauging of the R-symmetry. The bosonic field content of the theory consists of the metric tensor $g_{\mu\nu}$, $n_{\rm v}+1$ Abelian gauge fields $A_\mu^I$ and $n_{\rm v}+1$ scalars $X^I$. The scalars are subject to a constraint so that only $n_{\rm v}$ of them are independent, and are real in Lorentzian signature. The gauging of the R-symmetry is characterized by real constants $g_I$, the Fayet-Iliopoulos parameters, which give the charge of the gravitino under the gauge fields $A^I$; this means that the vector field gauging the R-symmetry is $g_I A^I$. 
 Up to boundary terms, the two-derivative bosonic action reads in $ (-++++)$ signature
\begin{equation}\label{eq:R-gauged_action}
S = \frac{1}{16\pi G}\int \Big[ 
\left(R - 2{\cal V}\right)\star 1 - Q_{IJ} \diff X^I \wedge \star \diff X^J - Q_{IJ} F^I \wedge \star F^J - \frac{1}{6}C_{IJK} F^I \wedge F^J \wedge A^K
\Big]\,,
\end{equation}
where $F^I=\diff A^I$, $I=1,2,\ldots,n_{\rm v}+1$, while $C_{IJK}$ is a completely symmetric constant tensor. The explicit form of the scalar potential $\mathcal{V}$ and the kinetic matrix $Q_{IJ}$ can be found e.g.\ in~\cite{Gutowski:2004yv} and will not be needed here.
 It will be convenient to redefine
\begin{equation}\label{eq:Ctok}
\frac{\pi}{4G}C_{IJK} \to k_{IJK}   \,,
\end{equation}
where $k_{IJK}$ is the cubic 't Hooft anomaly coefficient appearing in the previous sections.\footnote{The map relating $k_{IJK}$ and $C_{IJK}$  receives higher-derivative corrections in gauged supergravity~\cite{Cassani:2024tvk}.}

Supersymmetric solutions admit a globally defined Killing spinor, and their general local form can be determined by analyzing the conditions satisfied by the associated spinor bilinears~\cite{Gauntlett:2003fk,Gutowski:2004ez,Gutowski:2004yv,Gutowski:2005id,Bellorin:2006yr}. Among these bilinears one finds a Killing vector, denoted by $\xi$, which we shall refer to as the supersymmetric Killing vector.
We will focus on the  class of solutions known as timelike, where the supersymmetric Killing vector is timelike in Lorentzian signature. 
From now on we will work in Euclidean signature.\footnote{In particular, this means that our $\xi$ and the vielbein $e^0$ proportional to $\xi^\flat$ are related to the $\xi_{\rm L}$ and $e^0_{\rm L}$ used in Lorentzian signature by factors of $\ii$, $\xi = -\ii \xi_{\rm L} $ and $e^0 = \ii e^0_{\rm L}$.} 
The metric and gauge fields take the form  \cite{Gauntlett:2003fk,Gutowski:2004yv},
\be\label{eq:gen_form_sol}
\begin{aligned}
\diff s^2 \,&=\, (e^0)^2 + \diff s^2(\mathbb{B})\,,\\[1mm]
A^I \,&=\, -\ii X^Ie^0 - \eta^I\,,
\end{aligned}
\ee
where $\mathbb{B}$ is a K\"ahler base,   $e^0\, =\, \frac{\xi^\flat}{||\xi||}\,\equiv \, \frac{\xi^\mu g_{\mu\nu}  \diff x^\nu}{\left(g_{\rho\sigma} \xi^\rho\xi^\sigma\right)^{1/2}}$  describes how the orbits of the supersymmetric Killing vector $\xi$ fibre over the base, and $\eta^I $ are local one-forms defined on the base, up to gauge transformations. We refer to the original literature for the complete set of algebraic and differential conditions.

We are interested in complexified supersymmetric non-extremal solutions. More precisely, we consider Euclidean configurations that, at the two-derivative level, can locally be obtained by analytic continuation of the Lorentzian timelike solutions classified in~\cite{Gauntlett:2003fk,Gutowski:2004ez,Gutowski:2004yv,Gutowski:2005id,Bellorin:2006yr}. 
The recent analysis of~\cite{Colombo:2025ihp,Colombo:2025yqy} shows that, at least when U(1)$^3$ isometry is preserved, the global features of these solutions are characterized in terms of the chemical potentials specified asymptotically, together with some topological data from the metric and gauge fields. This allowed for the evaluation of their two-derivative on-shell action via equivariant localization, without requiring the explicit form of the metric and gauge fields, which in most cases remains unknown. In the following, we review the ingredients of this construction that will be needed in our analysis.

Note that higher-derivative corrections -- as long as they can be treated perturbatively -- may deform the solution but do not affect the existence of the supersymmetric Killing vector and the global data that we are going to introduce.

\subsection{Asymptotic boundary conditions}

We consider asymptotically AdS$_5$ solutions with compact Euclidean time.  Working in global coordinates, the metric asymptotically approaches
\begin{equation}
\label{eq:bdry_metric1}
\frac{\diff s^2}{\tilde{\ell}^2} \to \frac{\diff z^2}{z^2} + \frac{1}{z^2}\left[\frac{\beta^2}{4\pi^2}\diff \phi_0^2 + \diff \theta^2 + \sin^2\theta\left(\diff \phi_1+ \frac{\beta\Omega_1}{2\pi\ii}\diff\phi_0\right)^2 + \cos^2\theta\left(\diff\phi_2+\frac{\beta\Omega_2}{2\pi\ii}\diff\phi_0\right)^2 \right]\,,
\end{equation}
where  $\tilde{\ell}$ is the AdS radius and $\theta\in[0,\pi/2]$. 
 The angular coordinates obey the untwisted identifications
\begin{equation}
\label{eq:coord_id}
\left(\phi_0,\phi_1,\phi_2\right) \sim \left(\phi_0 + 2\pi,\phi_1,\phi_2\right) \sim \left(\phi_0,\phi_1+2\pi,\phi_2\right)\sim\left(\phi_0,\phi_1 ,\phi_2+2\pi\right)\,,
\end{equation}
which identify the basis of $2\pi$-periodic Killing vectors as
\begin{equation}
\label{eq:basisU1}
\partial_{\phi_0} \equiv \left(1,0,0\right) \,,\qquad \partial_{\phi_1} \equiv \left(0,1,0\right)  \,,\qquad \partial_{\phi_2} \equiv \left(0,0,1\right)  \,.
\end{equation}
Here, we introduced the shorthand notation 
\be
\left(a_0,a_1,a_2\right) \,\equiv\, a_0\partial_{\phi_0} + a_1 \partial_{\phi_1} + a_2\partial_{\phi_2}\,.
\ee
The conformal boundary at $z=0$ is, therefore, topologically a product of a round three-sphere $S^3$ and a circle, denoted by $S^1_{\rm bdry}$, of proper length $\beta$. The $S^3$ is fibered over the circle, with fibration terms controlled by the angular velocities $\Omega_1$, $\Omega_2$. 

We will denote by $\bar X^I$ the constant boundary value taken by the scalars $X^I$ in the two-derivative theory. These satisfy $g_I \bar X^I \propto \ell^{-1}$, where $\ell$ is the two-derivative AdS$_5$  radius.

In these conventions, the supersymmetric Killing vector $\xi$ may be expressed as
\begin{equation}
\label{eq:susyKV}
\xi \,=\, \frac{2\pi}{\beta}\partial_{\phi_0} + \frac{\ii\omega_1}{\beta}\partial_{\phi_1} + \frac{\ii\omega_2}{\beta}\partial_{\phi_2} \,=\, \frac{1}{\beta}\left(2\pi,\ii\omega_1,\ii\omega_2\right)\,,
\end{equation}
where we introduced the refined angular velocities~\cite{Silva:2006xv,Cabo-Bizet:2018ehj}
\begin{equation}
\omega_{1,2} \,=\, \beta\left(\Omega_{1,2}-\ell^{-1}\right)\,.
\end{equation} 

The boundary conditions also include holonomies of the gauge fields around $S_{\rm bdry}^1$, which define the electrostatic potentials
\begin{equation}
\label{eq:gauge_holonomy}
\varphi^I 
\,=\,  - \beta\bar X^I -\int_{S_{\rm bdry}^1} \!\!\!\! \ii A^I  \,.
\end{equation}

The condition that the asymptotic Killing spinor be well-defined 
(in particular, it must be antiperiodic around $S^1_{\rm bdry}$ since we want to allow for geometries where this circle is contractible in the bulk) 
 gives the supersymmetric constraint~\cite{Cabo-Bizet:2018ehj,Cassani:2025sim}
\begin{equation}
\label{eq:susy_constraint}
\omega_1 + \omega_2 - 2\varphi^{\cal R} \,=\, 2\pi \ii \qquad ({\rm mod} \ 4\pi\ii) \,,
\end{equation}
 where $\varphi^{\cal R}$ is the potential for the R-symmetry, under which the Killing spinor has charge 1. Using the Fayet-Iliopoulos gauging parameters $g_I$, this is expressed as 
\be
\varphi^{\cal R} \,=\, g_I \varphi^I\,.
\ee

By taking the limit $g_I = 0$ the theory reduces to ungauged supergravity. In this case, $\varphi^{\mathcal{R}}=0$ and the constraint  \eqref{eq:susy_constraint} simplifies to $\omega_1 + \omega_2 = 2\pi \ii \ ({\rm mod} \ 4\pi\ii)$. 


\subsection{Discrete data from the bulk}

\paragraph{Geometry of solutions with U(1)$^3$ symmetry.} All the relevant topological information of these geometries can be encoded in a collection of vectors $\{V^a\in \mathbb Z^3\}_{a=0,...,s}$, referred to as fan in analogy with toric geometries~\cite{Colombo:2025yqy}.\footnote{Similarly, for asymptotically flat solutions of five-dimensional supergravity all relevant topological information is encoded in a set of vectors that specify the \emph{rod structure}~\cite{Harmark:2004rm,Hollands:2007aj}. See~\cite{Breunholder:2017ubu} for a classification of asymptotically flat Lorentzian solutions based on the rod structure, and~\cite{Cassani:2025iix} for the extension to supersymmetric non-extremal complexified saddles.} Each $V^a$ specifies a U(1) subgroup of U(1)$^3$ that degenerates on a three-dimensional subspace ${\cal D}_a$, where the torus action has a U(1) isotropy group and that we may denote as bolt borrowing the terminology of~\cite{Gibbons:1979xm}.\footnote{We use the same notation to indicate the three-dimensional bolts $\mathcal{D}_a$ appearing here and their uplift to four-dimensional divisors in the extended six-dimensional geometry discussed in section~\ref{sec:equivariant_details}.} In the basis \eqref{eq:basisU1}, such vectors take the form
\begin{equation}
\label{eq:Va}
V^a = n_a\partial_{\phi_0} + r_a\partial_{\phi_1} + p_a\partial_{\phi_2} \,=\, \left(n_a,r_a,p_a\right)\,,
\end{equation}
where $n_a,r_a,p_a\in\mathbb Z$. Bolts with $n_a\neq 0$ are interpreted as Euclidean horizons, since a  vector  involving the Euclidean time vanishes there, while those with $n_a=0$ represent non-trivial topologies that can be present outside horizons -- also known as bubbles. 
Associated with each fixed locus ${\cal D}_a$ there is a normal space ${\cal N}_a$, invariant under the action of $V^a$ with the topology of a disk, obtained by foliating the orbits generated by $V^a$ over a radial direction transverse to ${\cal D}_a$. The normal space intersects the bolt once, and its boundary lies at asymptotic infinity, $\partial{\cal N}_a\subset \partial{\cal M}_5$. The intersection of two adjacent fixed loci ${\cal D}_a$ gives a fixed circle ${\cal D}_{a-1} \cap {\cal D}_a \cong S^1$ (referred to as nut in analogy with~\cite{Gibbons:1979xm}). The topology of the fixed loci ${\cal D}_a$ can be read directly from the fan. In particular, 
\begin{equation}
{\cal D}_a \cong L\left(\mathtt p,\mathtt q\right) \,,\qquad \mathtt p = \Big|\left[ V^{a-1},V^a,V^{a+1}\right]\Big|\,,
\end{equation}
while
\begin{equation}
\mathtt q = \left[ V^a,V^{a+1},w^a\right] \quad \text{mod}\,\, \mathtt p\,.
\end{equation}
Here, the vector $w^a\in \mathbb Z^3$ is chosen so that\footnote{Such vector is not unique, since it is defined up to shifts $w^a \to w^a + \mathbb Z\, V^{a-1} + \mathbb Z\, V^a$. Moreover, such vector does not exist when the corresponding locus has an orbifold singularity. In the following we focus, for simplicity, on five-dimensional geometries without orbifold singularities. 
}
\begin{equation}
\left[ V^{a-1} , V^a, w^a\right] =1 \,.
\end{equation}

The first and last vectors in the fan are always fixed by the asymptotic boundary metric \eqref{eq:bdry_metric1}. In our conventions they are
\begin{equation}
\label{eq:extreme_vectors}
V^0 = \left( 0,1,0\right) \,,\qquad V^s = \left( 0,0,1\right)\,,
\end{equation}
corresponding to the two circles that degenerate at the poles of the $S^3$ at the conformal boundary. They are therefore common to all asymptotically AdS$_5$ solutions with the same conformal boundary, including the reference Euclidean AdS$_5$ vacuum. In the latter geometry, the action of these two vectors smoothly degenerates at the center of space. 



\paragraph{Gauge field data.}

For geometries with non-trivial topology, the gauge fields $A^I$ are in general not globally defined. They must instead be defined patchwise, with gauge transformations relating the gauge fields on different patches.\footnote{For instance, we can introduce a convenient set of five-dimensional patches $\mathcal U_a$ following~\cite{Colombo:2025yqy}. The patch $\mathcal U_a$ intersects ${\cal D}_{a-1}$ and ${\cal D}_a$ (and contains their intersection), but no other ${\cal D}_{b}$, with $b\neq a-1,a$. The boundary of the patch ${\cal U}_a$ is a four-dimensional interface, $\partial {\cal U}_a = U_{a,a+1} - U_{a-1,a}$ (notice that $\partial {\cal U}_a \cap \partial {\cal U}_{a+1} \neq 0$ as well as $\partial {\cal U}_a \cap \partial {\cal U}_{a-1} \neq 0$). All the patches share a common corner, $\partial U_{a,a+1} = u_{0,...,s}$. The asymptotic boundary is covered by a single patch, ${\cal U}_0$, which intersects both ${\cal D}_0$ and ${\cal D}_s$.}   

On each patch we introduce a gauge field $A^I_a$, which is regular on that open set. The regularity conditions for the gauge field $A_a^I$ are: 
\begin{equation}
V^{a} \cdot A^I_a \Big|_{{\cal D}_a} \,=\, 0 \,=\, V^{a-1} \cdot A_a^I \Big|_{{\cal D}_{a-1}} \,.
\end{equation}
The gauge field on the boundary patch, $A^I= A_0^I$, is fixed by the choice of boundary condition \eqref{eq:gauge_holonomy}. For $a=1,...,s-1$, the gauge field $A_a^I$ differs from $A^I$ by a gauge transformation
\begin{equation}
\label{eq:gaugetransf}
A_a^I - A^I \,=\, \diff\zeta_a^I \,.
\end{equation}
Since ${\cal L}_V \diff\zeta_a^I =0$ for any vector $V$, we find it convenient to define the constants
\begin{equation}
\label{eq:qa}
V^a\cdot \diff\zeta_a^I \,\equiv\, q_a^I\,,
\end{equation}
with $a=0,1,...,s$. From this construction one immediately obtains that $q_0^I = 0 = q_s^I$. Moreover, using the regularity of $A^I_a$, one finds
\begin{equation}
\label{eq:qaI}
q_a^I \,=\, -V^a\cdot A^I\Big|_{{\cal D}_{a}}\,.
\end{equation}
%
These parameters need to satisfy suitable quantization conditions~\cite{Cassani:2025iix,Colombo:2025yqy} and characterize the gauge field holonomies around $\partial{\cal N}_a$, which can be translated into fluxes of the field strength across ${\cal N}_a$ as follows:\footnote{
To prove this relation one needs to use that
\begin{equation}
-\ii\int_{{\cal N}_a} F^I =  2\pi\ii V^a\cdot A^I\Big|_{{\cal D}_a} - n_a\int_{S_{\rm bdry}^1}\ii A^I = -2\pi \ii q_a^I - n_a\int_{S_{\rm bdry}^1}\ii A^I\,.
\end{equation}
} 
\begin{equation}
\label{eq:gauge_flux}
\begin{aligned}
\varphi^I_a &= -\int_{\partial{\cal N}_a} \left(\ii A^I_a + \beta \bar X^I \frac{\diff \phi_0}{2\pi}\right) = -\ii\int_{{\cal N}_a}F^I - \beta \bar X^I\int_{\partial{\cal N}_a}\frac{\diff\phi_0}{2\pi}
\\[1mm]
&= n_a \varphi^I - 2\pi \ii q_a^I\,.
\end{aligned}
\end{equation}

Thus, a potential $\varphi^I_a$ is associated with each bolt ${\cal D}_a$. For horizons (with $n_a\neq 0$) these potentials are proportional to the boundary conditions $\varphi^I$, up to a quantized shift, while for spatial bubbles ($n_a = 0$) they turn out to be proportional to the bubble potentials of~\cite{Cassani:2025iix}, which can be expressed as $q_a^I = \frac{1}{2\pi}\int_{{\cal N}_a} F^I 
$. Finally, using \eqref{eq:extreme_vectors} and \eqref{eq:qaI} one finds
\begin{equation}
\varphi^I_0 = 0 =\varphi^I_s\,.
\end{equation}

For each gauge field $A^I_a$ it is possible to introduce a suitable local one-form $\eta_a^I$,
\begin{equation}
\label{eq:eta}
\eta_a^I = - \ii X^I e^0 - A^I_a\,,
\end{equation}
where we recall that $ e^0\, =\, \frac{\xi^\flat}{||\xi||}$.
This one-form is regular on the corresponding bolt ${\cal D}_a$, whose differential is basic with respect to $\xi$, i.e. $\iota_\xi \diff \eta^I_a =0$. Then, their differentials $\diff \eta^I_a = \diff \eta^I$ (here $\eta^I$ denotes the one-form defined using the gauge fields $A^I$, as in \eqref{eq:gen_form_sol}) admit an expansion in cohomological classes of the four-dimensional K\"ahler base $\mathbb{B}$, which up to exact terms reads
\begin{equation}
\label{eq:deta}
\diff \eta^I \,=\, 2\pi \sum_{a=0}^s \alpha_a^I\,\mathtt c_1\big({\cal L}_a^{\mathbb B}\big)\,,
\end{equation}
where ${\cal L}_a^\mathbb B$ denotes the line bundle associated to ${\cal D}_a$ restricted to the base. Importantly, it has been proven in~\cite{Colombo:2025yqy} that the coefficients specifying such expansion are related to the potentials \eqref{eq:gauge_flux} by a simple map,
\begin{equation}
\label{eq:alphatophi}
\alpha_a^I \,=\, \frac{\varphi^I_a}{2\pi\ii}  \, =\, n_a\,\frac{ \varphi^I }{2\pi\ii} - q_a^I\,.
\end{equation} 
Hence the cohomology of $\diff\eta^I$ is controlled by the boundary conditions specified by the electrostatic potentials $\varphi^I$, together with the quantized data $n_a$, $q_a^I$  associated with each bolt.


\paragraph{Linear constraints from regularity.} The cohomological expansion of $\diff\eta$ \eqref{eq:deta} is constrained by the corresponding expansion of the Ricci form of the four-dimensional K\"ahler base. This implies a set of linear relations between the potentials \eqref{eq:alphatophi} and the integers characterizing the vector $V^a$ that vanishes at the bolt ${\cal D}_a$~\cite{Colombo:2025yqy}. These relations take the form
\begin{equation}
2\left[ \xi, V^0,V^s\right] \alpha_a^{\cal R} \,=\, \left[ \xi, V^0, V^a\right] + \left[ \xi, V^a, V^s\right] - \left[ \xi, V^0, V^s\right]\,,
\end{equation}
and may also be seen as relating the angular velocities of $V^a$, namely
\begin{equation}
\label{eq:omegaa}
\omega_{1}^a =  n_a\omega_{1} + 2\pi \ii r_a\,, \qquad \omega_{2}^a =  n_a\omega_{2} + 2\pi \ii p_a\,,
\end{equation}
to the potentials \eqref{eq:alphatophi},  as
\begin{equation}
\label{eq:susy_constraint2}
\omega_1^a + \omega_2^a - 2\varphi^{\cal R}_a \,=\, 2\pi \ii\,.
\end{equation}
A consequence of these linear relations is that the Killing spinor is antiperiodic around the contractible orbits generated by $V^a$, as required by its regularity. 
Using the boundary constraint \eqref{eq:susy_constraint} (with right-hand side fixed to $2\pi \ii$), this reduces to a condition on the integers,
\begin{equation}
\label{eq:int_constraint}
n_a + r_a + p_a + 2q_a^{\cal R} = 1\,,\qquad\qquad \text{with} \ \ q_a^{\cal R}=g_I q_a^I  \,.
\end{equation}
In the gauged case this condition may be solved for $q_a^{\cal R}$ in terms of the integers defining the vector  $V^a$. 
In the ungauged limit the Killing spinor becomes uncharged under all ${\rm U}(1)_I$ gauge symmetries in the theory, hence $q_a^{\cal R}=0$. Then \eqref{eq:int_constraint} reduces to a constraint  on the vector $V^a$ itself,  which may be solved as $r_a = 1-n_a-p_a$~\cite{Cassani:2025iix}.

\section{Black holes, rings and lenses from anomaly polynomial}
\label{sec:OSAfromAP}

\subsection{General formula for the on-shell action}\label{general_formula_action}

We can now give our main result, namely a general formula for the on-shell action of supersymmetric solutions with U(1)$^3$ symmetry of the type described in section~\ref{sec:sugra_sol}, including higher-derivative corrections. 

This is obtained by putting together the ingredients introduced in the previous sections. First, we note that the on-shell action \eqref{eq:CS_HD} takes the same form as the anomaly polynomial~\eqref{eq:P}, with $F^I$ being replaced by the form $-\diff\eta^I =  \diff A^I + \ii\, \diff(X^I \,e^0)$ defined in \eqref{eq:eta}. 
Of course, this assumes that our conjectural rewriting of the higher-derivative part of the on-shell action given in \eqref{eq:CS_HD} is correct.
Also we should use a suitable six-dimensional extension of the local one-forms $\eta^I$, together with the corresponding cohomological expansion of their differentials \eqref{eq:deta}.\footnote{The relevant expansion in Chern classes to be used in the localization formula is then the one given in \eqref{eq:deta}, where the Chern classes of the K\"ahler base should now be understood as extended to $\mathcal{M}_6$. It would be desirable to formulate this extension more rigorously. Note that we use the same symbol $\eta$ to denote the original connection defined on $\mathcal{M}_5$ and its extension to $\mathcal{M}_6$. While the former satisfies $\iota_\xi\diff\eta=0$, the latter fails to satisfy this relation because of the dependence on the additional coordinate. We hope this does not cause confusion.}

Then we can evaluate the action integral~\eqref{eq:CS_HD} via equivariant localization, as explained in section~\ref{sec:equivariant_details}. 
We localize with respect to the action of the supersymmetric Killing vector $\xi$, expressed as in \eqref{eq:susyKV}. The equivariantization of the action \eqref{eq:CS_HD} is easily obtained by replacing $\diff\eta$ appearing there with $\diff_\xi\eta=(\diff+\iota_\xi)\eta$.

Then the supersymmetric on-shell action, including higher-derivative corrections, is computed by:
\be
I \,=\,  2\pi \ii \int_{\mathcal M_6}\mathtt P \,,
\ee
with the right hand side given by the localized formula~\eqref{eq:int_an_poly}--\eqref{eq:int_an_poly_4},
where one should implement the replacement
\begin{equation}\label{eq:replacement}
\DD_{k_i}^I \,\to\, \frac{\varphi^I_{k_i}}{2\pi\ii} \,=\, n_{k_i}\,\frac{\varphi^I}{2\pi\ii} - q_{k_i}^I \,,
\end{equation}
as specified by the map \eqref{eq:alphatophi}. 

We thus arrive at the following general formula for the on-shell action:
\begin{equation}
\label{eq:result_action}
I \,=\,  \sum_{\CC_k}\left(\frac{1}{6}\,k_{IJK} I^{IJK}(\CC_k) - \frac{1}{24}\,k_J I^J(\CC_k)\right)\,,
\end{equation}
where the contribution of each simplicial cone $\CC_k$ arising from the decomposition of the fan is
\begin{equation}\label{eq:result_action_2}
I^{IJK}(\CC_k)= \frac{\big(\sum_{i=1}^3 \epsilon_{k_i}(n_{k_i}\varphi^I-2\pi\ii q_{k_i}^I)\big)\,\big(\sum_{i=1}^3 \epsilon_{k_i}(n_{k_i}\varphi^J-2\pi\ii q_{k_i}^J)\big)\,\big(\sum_{i=1}^3 \epsilon_{k_i}(n_{k_i}\varphi^K-2\pi\ii q_{k_i}^K)\big)}{ (2\pi)^2 \,  d_k\,\epsilon_{k_1}\epsilon_{k_2}\epsilon_{k_3}},
\end{equation}
together with
\begin{equation}\label{eq:result_action_4}
I^J(\CC_k)\,=\,  -\frac{\big(\sum_{i=1}^3 \epsilon_{k_i}(n_{k_i}\varphi^J-2\pi\ii q_{k_i}^J)\big)\left[(\epsilon_{k_1})^2+(\epsilon_{k_2})^2+(\epsilon_{k_3})^2\right]}{  d_k\,\epsilon_{k_1}\epsilon_{k_2}\epsilon_{k_3}}\,.
\end{equation}
We recall that the $\epsilon_{k_{1,2,3}}$ are the weights of $\xi$ in the basis of vectors that specifies the cone $\CC_k$, and are determined by \eqref{eq:equiv_params}.
The chemical potentials $\omega_1,\omega_2,\varphi^I$ specified at the boundary need to satisfy the linear constraint \eqref{eq:susy_constraint}, and  one must also impose the integer conditions \eqref{eq:int_constraint}.

The first piece in \eqref{eq:result_action} is the two-derivative term, where only the coupling $k_{IJK}$ may receive higher-derivative corrections, while the second piece is the genuine higher-derivative term.

We emphasize that the formula holds both in gauged and ungauged supergravity, with the only differences appearing in the constraints to be imposed.

In the following we apply our general formula to some examples of interest.

\subsection{Black hole}

The supersymmetric five-dimensional black hole with spherical horizon topology is a geometry in which the $S^1$ of the boundary metric degenerates in the bulk at a locus ${\cal D}_1\cong S^3$, giving rise to a disk topology ${\cal N}_1$. There are no further non-trivial topological features. In our conventions, this geometry is described by the fan~\cite{Cabo-Bizet:2018ehj,Cassani:2024kjn,Colombo:2025ihp}
\begin{equation}
\label{eq:BHfan}
V^0 = \left( 0,1,0\right)\,,\qquad V^1 = \left(1,0,0\right) \,,\qquad V^2 = \left( 0,0,1\right)\,.
\end{equation}
The regularity constraint \eqref{eq:int_constraint} is trivially satisfied with $q_1^{\cal R}=0$. For simplicity, we take $q_1^I =0$, which is the case admitting a smooth limit to minimal supergravity.

\paragraph{Six-dimensional extension and on-shell action.}

A smooth six-dimensional extension of this spacetime, providing a suitable background for the equivariant integration of the anomaly polynomial, was already considered in~\cite{Cassani:2024tvk}. In that construction, the six-dimensional space has topology ${\cal M}_6 = {\cal M}_4 \times D_2$, where ${\cal M}_4$ forms a cone over $S^3$ and $D^2$ is a disk, with $\partial D_2 = S^1$. In this extension the torus action has a unique fixed point at the origin of space. The constraint \eqref{eq:susy_constraint} must still be imposed so that the Killing spinor is well-defined on the cigar geometry. 

Indeed, the vectors \eqref{eq:BHfan} define the fan of a six-dimensional cone. The Calabi-Yau condition is satisfied, since, after a suitable ${\rm SL}\left(3,\mathbb Z\right)$ rotation, the vectors can be written as
\begin{equation}
V^0 \to \tilde V^0 = \left(1,0,0\right)\,,\qquad V^1 \to \tilde V^1 = \left(1,-1,0\right)\,,\qquad V^2 \to \tilde V^2 = \left(1,0,1\right)\,.
\end{equation}
Since exactly three facets meet at its tip, no triangulation is required. The local geometry at the tip of the cone is $\mathbb C^3$, and each vector in the fan rotates a copy of $\mathbb C \subset \mathbb C^3$. The weights of the torus action are specified by the corresponding equivariant parameters
\begin{equation}
\label{eq:equiv_pars_BH}
\epsilon_0 = \frac{\left[ \xi,V^1,V^2\right]}{\left[V^0,V^1,V^2\right]} = \frac{\ii\omega_1}{\beta}\,,\qquad \epsilon_1 =\frac{ \left[ V^0,\xi,V^2\right]}{\left[V^0,V^1,V^2\right]} = \frac{2\pi}{\beta}\,,\qquad \epsilon_2 = \frac{\left[V^0,V^1,\xi\right]}{\left[V^0,V^1,V^2\right]} = \frac{\ii\omega_2}{\beta}\,.
\end{equation}
The gauge data that specify the Chern classes \eqref{eq:gauge_chern} are chosen to reproduce the fluxes \eqref{eq:gauge_flux} of the supergravity solution, that is
\begin{equation}
\DD^I_0 = 0 = \DD^I_2 \,,\qquad \DD_1^I = \frac{\varphi^I}{2\pi\ii}\,.
\end{equation}
The general formula \eqref{eq:result_action} for the on-shell action then reduces to
\begin{equation}\label{eq:blackholeaction}
\II \,=\, \frac{k_{IJK}\varphi^I\varphi^J\varphi^K}{6\,\omega_1\omega_2} - \frac{k_I\varphi^I\left(\omega_1^2 + \omega_2^2 - 4\pi^2\right)}{24\,\omega_1\omega_2}\,,
\end{equation}
which reproduces the known results for the black hole on-shell action, including higher-derivative corrections, as already discussed in~\cite{Cassani:2022lrk,Bobev:2022bjm,Cassani:2024tvk}.\footnote{Recently, this result has also been reproduced by localizing the five-dimensional action, including a certain class of higher-derivative corrections, using equivariant localization in five-dimensional supergravity~\cite{BenettiGenolini:2026qdm,Gaar:2026nqq}.} 

 In the asymptotically flat case, the Legendre transform of \eqref{eq:blackholeaction}, yielding the Wald entropy of the BMPV black hole, has been found in perfect agreement with a microscopic index computation, including the higher-derivative corrections~\cite{Alexandrov:2026rra}.

In the asymptotically AdS$_5$ case, this expression has been shown to precisely reproduce the Cardy-like asymptotics of the superconformal index of $\mathcal{N}=1$ SCFT's with flavour chemical potentials turned on~\cite{Cassani:2024tvk}. 
 Based on this, we expect that \eqref{eq:blackholeaction} also gives the correct action for asymptotically AdS$_5$ black hole saddles that are solutions of 10d supergravity on some internal manifold $Y_5$ (or 11d supergravity on $Y_6$), even in setups where five-dimensional Fayet-Iliopoulos gauged supergravity does not arise as a consistent truncation. For instance, by choosing suitable $k_{IJK}$ and $k_I$, eq.~\eqref{eq:blackholeaction} should describe the on-shell action of a supersymmetric AdS$_5$ black hole in type IIB supergravity on $T^{1,1}$ with the R-charge, the baryonic charge, and both flavour charges turned on, which has not been constructed so far.


\subsection{A discrete family from shifts and orbifolds}

A discrete family of Euclidean geometries which are locally equivalent to the black hole, but differ globally from it, can be obtained by performing a suitable $\mathbb Z_{|n|}$ orbifold of the black hole geometry. In the gauged case, the action of the orbifold also involves the ${\rm U}(1)_R$-symmetry (namely, when the solution is uplifted to ten or eleven-dimensional supergravity, the orbifold also acts on the isometry in the internal manifold corresponding to the dual SCFT R-symmetry~\cite{Aharony:2021zkr}).
The fan for this class of five-dimensional geometries is~\cite{Cassani:2025iix,Colombo:2025yqy} \footnote{The orbifold action can be described as follows (see~\cite{Aharony:2021zkr} for more details). We start from a seed black hole solution with fan \eqref{eq:BHfan}, and denote the corresponding fields and coordinates by a superscript ``tilde''. The orbifold acts by imposing an additional coordinate identification together with an R-symmetry gauge transformation,
\begin{equation}
\left(\tilde \phi_0,\tilde\phi_1,\tilde\phi_2\right)\sim\left(\tilde \phi_0+\frac{2\pi}{n},\tilde \phi_1-\frac{2\pi r}{n},\tilde \phi_2-\frac{2\pi p}{n}\right)\,,\qquad g_I \tilde A^I  \longrightarrow g_I A^I = g_I \tilde A^I - g_I q^I \diff \tilde \phi_0\,.
\end{equation}
Here $(\tilde\phi_0,\tilde\phi_1,\tilde\phi_2)$ are angular coordinates such that $V^0=\partial_{\tilde\phi_1}$, $V^1=\partial_{\tilde\phi_0}$ and $V^2=\partial_{\tilde\phi_2}$. 
Locally, the orbifold does not modify the bulk solution. However, although the seed connections $\tilde A^I$ are regular at the horizon, the shifted connections $A^I$ need not be regular there. After performing a suitable coordinate transformation $(\tilde\phi_0,\tilde\phi_1,\tilde\phi_2)\mapsto(\phi_0,\phi_1,\phi_2)$, where the new coordinates obey the identifications \eqref{eq:coord_id}, the seed fan takes precisely the form of the orbifold fan \eqref{eq:fan_orbifold}. In the same variables, the shift from $\tilde A^I$ to $A^I$ takes the form of a gauge transformation \eqref{eq:gaugetransf}, with $\tilde A^I$ being the connection regular at the bolt. The resulting orbifold may be freely acting in ten dimensions, where the gauge transformations are geometrized into rotation in the internal space; in particular, this happens if all parameters $n,r,p,q^I$ are non-zero. It can also have fixed points, with the precise type of singularity depending also on the topology of the internal manifold.
}
\begin{equation}
\label{eq:fan_orbifold}
V^0 = \left(0,1,0\right)\,,\qquad V^1 = \left(n,r,p\right)\,,\qquad V^2 = \left(0,0,1\right)\,.
\end{equation}
The fixed locus of the Killing vector $V^1$ is a Euclidean horizon with topology ${\cal D}_1 \cong S^3/\mathbb Z_{|n|}$. The regularity condition for the Killing spinor \eqref{eq:int_constraint} therefore imposes the constraint~\cite{Aharony:2021zkr}
\begin{equation}
\label{eq:constraint_orbifold}
n+r+p+2g_I q^I = 1\,.
\end{equation}

\paragraph{Six-dimensional extension and on-shell action.} In the six-dimensional extension with fan \eqref{eq:fan_orbifold},\footnote{Note that this geometry satisfies the Calabi-Yau condition only provided $p+r = 1$ mod $n$. Under this condition, one can find the two-dimensional vectors
$$
v^0 = \left(0,0\right)\,\qquad v^1 = \left(p,n\right)\,,\qquad v^2 = \left(1,0\right)\,,
$$
which for $n\neq 0$ form a triangle.} the torus action has equivariant parameters
\begin{equation}
\epsilon_0 \,=\,\ii\, \frac{n\omega_1 + 2\pi \ii r}{n\beta}\,,\qquad \epsilon_1 \,=\, \frac{2\pi}{n\beta}\,,\qquad \epsilon_2 = \ii\,\frac{n\omega_2 + 2\pi \ii p}{n\beta}\,.
\end{equation}
As above, the gauge data are chosen to reproduce the supergravity potentials \eqref{eq:gauge_flux}
\begin{equation}
\DD_0^I \,=\, 0 \,=\, \DD_2^I \,,\qquad \DD_1^I \,=\, \frac{n\varphi^I - 2\pi \ii q^I}{2\pi\ii}\,.
\end{equation}
Our formula for the action then gives
\begin{equation}
\begin{aligned}
\II \,& =\, \frac{k_{IJK}}{6}\frac{\left(n\varphi^I-2\pi \ii q^I\right)\left(n\varphi^J-2\pi \ii q^J\right)\left(n\varphi^K-2\pi \ii q^K\right)}{n\left(n\omega_1+2\pi \ii r\right)\left(n\omega_2+ 2\pi \ii p\right)} 
\\[1mm]
\,&\,\quad - \frac{k_I}{24}\,\frac{\left(n\varphi^I-2\pi \ii q^I\right) \left [\left( n\omega_1 + 2\pi \ii r\right)^2 + \left( n\omega_2 + 2\pi \ii p\right)^2  - 4\pi^2\right]   }{n\left(n\omega_1+2\pi \ii r\right)\left(n\omega_2+ 2\pi \ii p\right)}\,,
\end{aligned}
\end{equation}
where $n,r,p,q^I$ are subject to the constraint \eqref{eq:constraint_orbifold}.
 The first line precisely reproduces the action of the orbifold black hole saddles of~\cite{Aharony:2021zkr}. The second line is a prediction for the higher-derivative corrections to the action of such saddles. The whole expression is the expected action obtained by orbifolding a black hole with action~\eqref{eq:blackholeaction}; indeed orbifolding amounts to implementing the substitutions $\omega_i \to n\omega_i +2\pi\ii r$, $\varphi^I \to n\varphi^I-2\pi \ii q^I$ and dividing the action by $n$~\cite{Aharony:2021zkr}.\footnote{A careful patchwise treatment of the action integral led the authors of~\cite{Colombo:2025yqy} to add a purely imaginary term independent of the chemical potentials and proportional to $k_{IJK}q^Iq^Jq^K$ to the expression for the two-derivative on-shell action (cf.\ section 5.1.1 therein). It would be interesting to clarify if this can also be obtained using our approach.}
 The $k_I$ coefficients vanish for type IIB supergravity on $S^5$, but are generally non-zero when a different internal manifold is considered, see~\cite{Cassani:2024tvk} for an explicit example.


\subsection{Black lenses}

Euclidean saddles with lens space horizon $L(2,1)$ and a single bubble outside of it are characterized by two possible choices of fan, characterized by a sign choice, $\sigma =\pm 1$~\cite{Cassani:2025iix,Colombo:2025yqy}:
\begin{equation}
\label{eq:lens_fan}
V^0=\left(0,1,0\right)\,,
\qquad
V^1=\left(0,2,-\sigma\right)\,,
\qquad
V^2=\left(1,0,0\right)\,,
\qquad
V^3=\left(0,0,1\right)\,.
\end{equation}
 In both cases, the compact fixed loci associated with the Killing vectors $V^1$ and $V^2$ are, respectively, a bubble ${\cal D}_1\cong S^3$ and the lens space horizon ${\cal D}_2\cong L(2,1)$. 
%

We need to impose the regularity condition \eqref{eq:int_constraint} in each compact bolt. For the horizon, this condition is solved by taking $q_2^{\cal R} =0$. For simplicity, we set $q_2^I =0$. The regularity condition associated with the bubble ${\cal D}_1$ instead gives
\begin{equation}
\label{eq:q1_BL}
q_1^{\cal R} = \frac{\sigma - 1}{2}\,.
\end{equation}
This is the first difference between the two choices of fan:
for $\sigma =1$, one has $q_1^{\cal R} =0$, 
while for $\sigma =-1$ one obtains $q_1^{\cal R} =-1$. It follows that, while the former condition cannot be solved in minimal supergravity (where $q_1^I = q_1$ for all values of the index ``$I$''), unless one also takes the ungauged limit, the latter case does not admit a limit to ungauged supergravity, since the corresponding constraint cannot be solved when $g_I=0$. 

We turn on non-zero bubble potentials supported by the non-trivial topology, i.e. $q_1^I \neq 0$ (while setting $q_2^I = 0$). We therefore take
\begin{equation}
\label{eq:delta_lens}
\DD_1^I = - q_1^I\,,
\qquad
\DD_2^I = \frac{\varphi^I}{2\pi\ii}\,,
\end{equation}
with $q_1^I$ constrained by \eqref{eq:q1_BL}. In the following, we drop the subscript ``1'' in $q_1^I$, i.e.\ we use $q^I\equiv q_1^I$.

\paragraph{Six-dimensional extension and on-shell action.} 

Next, we consider the six-dimensional extension of the five-dimensional geometry with fan \eqref{eq:lens_fan}. Note that for $\sigma =-1$ the fan is non-convex, therefore the six-dimensional extension does not define a genuine toric geometry. We will nevertheless apply the localization formula over this space, extracting the required fixed-point data by extrapolation from a convex case.\footnote{A closely related approach was used in~\cite{Martelli:2023oqk,Colombo:2023fhu} to treat geometries involving fibrations over a spindle. In those examples, the toric data depend on a parameter denoted by $\sigma$, with $\sigma =1$ corresponding to the \emph{twist} case, while $\sigma =-1$ to the \emph{anti-twist}. In the latter case the fan is non-convex. Nevertheless, the fixed-point data are determined by first considering a convex case, and then formally continuing the results to $\sigma \to -1$.}

\begin{figure}
	\centering
	\includegraphics[width=0.3\textwidth]{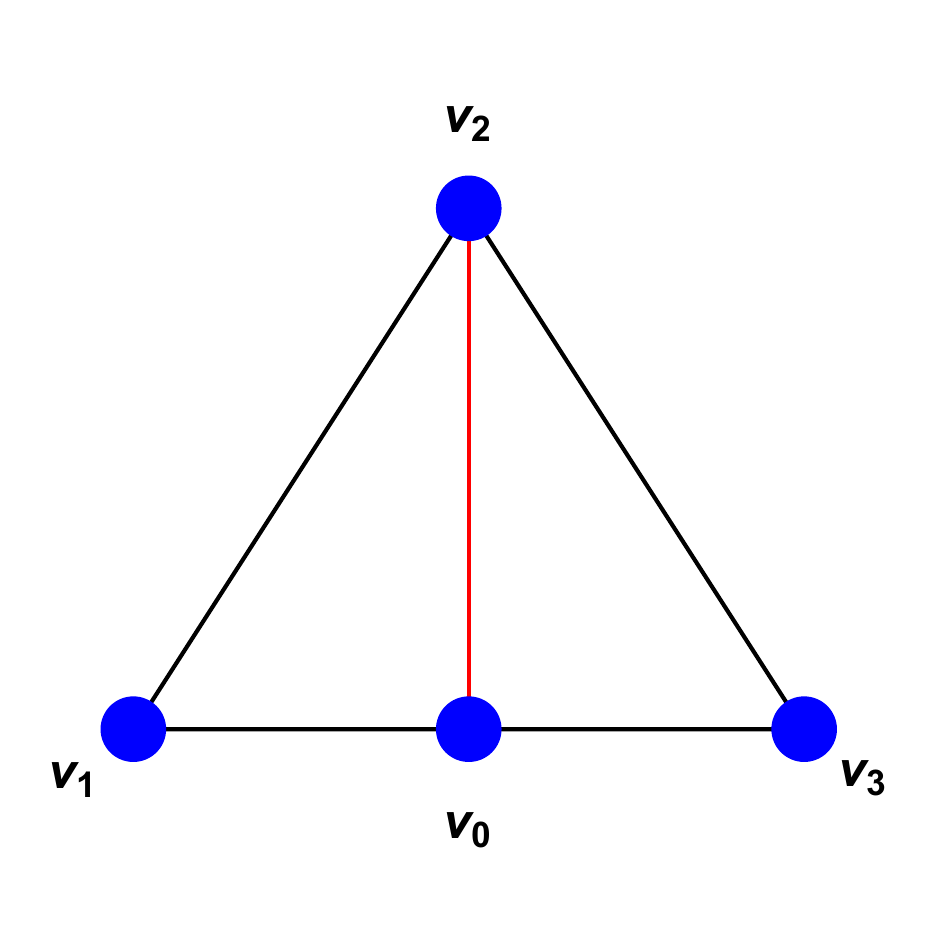}
	\caption{\it  Toric diagram for the black lens with $\sigma =1$. The diagonal $v^0-v^2$ specifies the triangulation used in the resolution.}
	\label{fig:diagram_lens}
\end{figure}

Let us first then discuss the case $\sigma=1$. In this case the fan is convex and satisfies the Calabi-Yau condition. Indeed, by means of a suitable $\mathrm{SL}(3,\mathbb Z)$ transformation, the vectors can be brought to the form $V^a \to \tilde V^a=(1,v^a)$, with
\begin{equation}
v^0 = \left(0,0\right)\,,\qquad v^1 = \left(-1,0\right)\,,\qquad v^2 = \left(0,1\right)\,,\qquad v^3 = \left( 1,0\right)\,.
\end{equation}
Before resolution, the corresponding cone is singular. In order to resolve the singularity, we triangulate the fan as illustrated in Fig.~\ref{fig:diagram_lens}. 
The relevant triangulation is obtained by adding the diagonal $v^0-v^2$, which gives the two cones (one for each compact bolt of the original five-dimensional geometry)
\begin{equation}
\label{eq:cones_BL}
\CC_1 = \{ V^0, V^1, V^2\}\,,\qquad \CC_2 = \{ V^0,V^2,V^3\}\,,
\end{equation}
each one corresponding to a fixed point of the torus action. Moving forward, we use the same pair of cones to extract the fixed-point data for generic $\sigma$:\footnote{If one formally continues the fan to generic values of $\sigma$, while remaining in a convex geometry, the Calabi-Yau condition is generically lost. In that case the fixed-point data should be extracted from a refinement obtained by studying the fan directly in three dimensions. With this prescription, the relevant cones are still those in \eqref{eq:cones_BL}, since $V^0$, $V^1$ and $V^3$ are coplanar and therefore cannot define a three-dimensional cone.} while for $\sigma =1$ this corresponds to the standard triangulation explained above, for $\sigma = -1$ this should be understood as an extrapolation of the localization data from the convex case. 

Keeping $\sigma$ arbitrary, the equivariant parameters associated with the first cone $\CC_1$ are
\begin{equation}
\epsilon_0 =\frac{\left[\xi, V^1, V^2\right]}{\left[V^0,V^1,V^2\right]} = \ii \frac{\omega_1 + 2\sigma^{-1}\omega_2}{\beta}\,,\quad \epsilon_1 =\frac{\left[V^0,\xi,V^2\right]}{\left[V^0,V^1,V^2\right]}= -\ii\frac{\sigma^{-1}\omega_2}{\beta}\,,\quad \epsilon_2 = \frac{\left[V^0,V^1,\xi\right]}{\left[V^0,V^1,V^2\right]} = \frac{2\pi}{\beta} .
\end{equation}
These weights encode a lens space topology and depend explicitly on the choice of $\sigma$. The equivariant parameters associated with the second cone $\CC_2$ coincide with those of the black hole cone in \eqref{eq:equiv_pars_BH},
\begin{equation}
\epsilon_0 \,=\, \ii\frac{\omega_1}{\beta}\,,\qquad \epsilon_1 \,=\, \frac{2\pi }{\beta}\,,\qquad \epsilon_2 \,=\, \ii\frac{\omega_2}{\beta}\,.
\end{equation}
Then, the general formula \eqref{eq:result_action}, with $\DD_a^I$ given by \eqref{eq:delta_lens} and $\DD_0^I = 0= \DD_3^I$, gives the on-shell action of the candidate asymptotically AdS$_5$ black lenses, including the higher-derivative corrections controlled by $k_I$,
\begin{equation}
\label{eq:action_BL}
\begin{aligned}
\II \,&=\, \frac{k_{IJK}}{6}\Bigl[\frac{\varphi^I\varphi^J\varphi^K}{\omega_1\omega_2}- \frac{\left(\varphi^I - \sigma^{-1}\omega_2q^I\right)\left(\varphi^J - \sigma^{-1}\omega_2q^J\right)\left(\varphi^K - \sigma^{-1}\omega_2q^K\right)}{\omega_2\left(\omega_1 + 2\sigma^{-1}\omega_2\right)} \Bigr]
\\[1mm]
\,&\,\quad - \frac{k_I}{24}\Bigl[\frac{\varphi^I\left(\omega_1^2 + \omega_2^2 -4\pi^2\right)}{\omega_1\omega_2}
- \frac{\left(\varphi^I -\sigma^{-1} \omega_2q^I\right)\left(\omega_1^2 + 4\sigma^{-1}\omega_1\omega_2 + 5\sigma^{-2}\omega_2^2 -4\pi^2\right)}{\omega_2 \left(\omega_1 + 2\sigma^{-1}\omega_2\right)}
\Bigr]\,.
\end{aligned}
\end{equation}
The first line reproduces known two-derivative results for both $\sigma = \pm 1$: for $\sigma =1$ it agrees with the two-derivative computation of~\cite{Cassani:2025iix,Colombo:2025yqy}, while for $\sigma =-1$ it matches the result in~\cite{Colombo:2025yqy}. 

Moreover, while the putative AdS$_5$ black lens with $\sigma =-1$ does not allow for an ungauged limit $g_I=0$, the $\sigma =1$ case does. In this case,   \eqref{eq:action_BL} provides a prediction for the higher-derivative corrections to the action and entropy of asymptotically flat supersymmetric black lenses, which are explicitly known. Indeed, Euclidean supersymmetric solutions with fan \eqref{eq:lens_fan} were recently constructed in~\cite{Cassani:2025iix,Boruch:2025sie} in ungauged (two-derivative) supergravity. 
These provide supersymmetric non-extremal generalizations of the Lorentzian BPS black lens of~\cite{Kunduri:2014kja,Kunduri:2016xbo} and should contribute as saddles to the flat-space gravitational index. 
It would be interesting to compare \eqref{eq:action_BL} (or the corresponding entropy obtained by Legendre transform) 
against a microscopic realization of such index, possibly obtained via the embedding in string theory discussed in \cite{Kunduri:2016xbo}.


\subsection{Black rings}\label{sec:black_rings}

Supersymmetric black ring saddles are characterized by two possible fans~\cite{Colombo:2025yqy}\footnote{We thank the authors of~\cite{Colombo:2025yqy} for drawing our attention to the possibility of a fan with $\sigma=-1$, which is discussed in the revised version of their work. }
\begin{equation}
\label{eq:fan_BR}
V^0 = \left(0,1,0\right) \,,\qquad V^1 = \left(0,0,\sigma\right)\,,\qquad V^2 = \left(1,0,0\right)\,,\qquad V^3 = \left(0,0,1\right)\,,
\end{equation}
where $\sigma =\pm 1$ is again a sign choice. 
For both values of $\sigma$, the fixed loci of $V^1$ and $V^2$ are the bubble ${\cal D}_1 \cong S^3$ and the horizon ${\cal D}_2 \cong S^1 \times S^2$, respectively. In these conventions, rotations around the axis of $S^2$ are generated 
by $\partial_{\phi_2}$.

For the fan \eqref{eq:fan_BR}, the regularity constraints \eqref{eq:int_constraint} are solved by
\begin{equation}
\label{eq:BR_constraint}
q^{\cal R}_1 = \frac{1-\sigma}{2}\,,\qquad q^{\cal R}_2=0\,.
\end{equation}
As in the black lens example, we take $q_2^I=0$ and $q_1^I \equiv q^I\neq 0$. 

Similarly to the black lens, the existence of a limit to an asymptotically flat black ring depends on the choice of $\sigma$. For $\sigma=1$, the condition \eqref{eq:BR_constraint} is solved by $q^{\cal R} =0$, which is compatible with the ungauged limit. Indeed, a two-derivative asymptotically $S^1\times \mathbb{R}^4$ solution with toric symmetry was constructed in~\cite{Cassani:2025iix}, providing a supersymmetric non-extremal generalization of the BPS black ring of~\cite{Elvang:2004rt,Elvang:2004ds}. See also \cite{Boruch:2025sie} for a construction with just ${\rm U}(1)^2$ isometry, and \cite{Dharanipragada:2026dji} for the construction of a small black ring saddle.

For $\sigma =-1$ the constraint reduces to $g_I q^I = 1$, therefore the corresponding solution can only belong to the gauged theory. 
Although an asymptotically AdS$_5$ black ring has not been found to date,\footnote{The near-horizon analysis of~\cite{Kunduri:2006uh,Grover:2013hja} excludes smooth supersymmetric black rings in minimal gauged supergravity, while allowing for their possible existence once couplings to vector multiplets are turned on~\cite{Kunduri:2007qy}.} 
we note that this relation agrees with a constraint obtained in the near-horizon analysis of~\cite{Kunduri:2007qy}, where a smooth near-horizon geometry with ring horizon topology was discussed. A detailed comparison with the near-horizon solution of~\cite{Kunduri:2007qy} will be presented in section~\ref{sec:compare_near_horizon}.

\paragraph{Six-dimensional extension and on-shell action.}

We discuss now the six-dimensional extension for the geometries with fan \eqref{eq:fan_BR} and present the result for the integral of the anomaly polynomial over these backgrounds. 

For $\sigma=1$, the fan \eqref{eq:fan_BR} satisfies the Calabi-Yau condition. In this case the fan can be brought to the form $V^a \to \tilde V^a = \left(1,v^a\right)$ with\footnote{The fan of the $\sigma =1$ black ring is degenerate, since $v^1=v^3$. However, we can still use it to extract the relevant triangulation data by a formal analytic continuation, similarly to the previous case of the non-convex black lens. To do so, we deform the fan to remove the degeneracy by introducing a toric diagram with vertices $\left(0,0\right)$, $\left(p,n\right)$, $\left(0,1\right)$ and $\left(1,0\right)$ (ordered clockwise) so as to form a convex quadrilateral. The original fan is then recovered by continuation in the limit $n\to 0$ and $p\to 1$. Among the two possible triangulations, we choose the one that remains non-singular in this limit. This prescription leads to the cones \eqref{eq:BR_cones}.}
\begin{equation}
v^0 = \left(0,0\right)\,,\qquad v^1 = \left(1,0\right)\,,\qquad v^2 = \left(0,1\right) \,,\qquad v^3 = \left(1,0\right)\,.
\end{equation}
Instead, for $\sigma =-1$ (or for general values of $\sigma$ obtained by a formal analytic continuation) one should determine a refinement by considering directly the three-dimensional fan. 
Then, the relevant toric decomposition identifies the two simplicial cones
\begin{equation}
\label{eq:BR_cones}
\CC_1 = \{ V^0, V^1, V^2\}\quad \text{and}\quad \CC_2 = \{V^0,V^2,V^3\}\,.
\end{equation}
Keeping $\sigma$ arbitrary, the equivariant parameters associated with the first cone are
\begin{equation}
\epsilon_0 \,=\, \ii\frac{\omega_1}{\beta}\,,\qquad \epsilon_1 \,=\, \ii \frac{\sigma^{-1}\omega_2}{\beta}\,,\qquad \epsilon_2 \,=\, \frac{2\pi}{\beta} \,,
\end{equation}
while those for the second cone coincide with the black hole ones \eqref{eq:equiv_pars_BH}, 
\begin{equation}
\epsilon_0 \,=\, \ii\frac{\omega_1}{\beta}\,,\qquad \epsilon_1 \,=\, \frac{2\pi}{\beta}\,,\qquad \epsilon_2 \,=\, \ii \frac{\omega_2}{\beta}\,.
\end{equation}
The gauge data are chosen as in the black lens case: 
\begin{equation}
\DD^I_0=0\,=\, \DD^I_3 \,,\qquad \DD^I_1 \,=\, -q^I\,,\qquad \DD^I_2 \,=\, \frac{\varphi^I}{2\pi\ii}\,.
\end{equation}

Our general formula~\eqref{eq:result_action} then gives a prediction for the on-shell action of the putative asymptotically AdS$_5$ black rings with $\sigma =\pm1$, as well as for the asymptotically flat black ring saddle. The integral localizes to a sum over the tips of the two cones \eqref{eq:BR_cones}. The second cone gives precisely the black hole action~\eqref{eq:blackholeaction}, while the first cone contributes with minus the same expression, where each electrostatic potential $\varphi^I$ is replaced by $\varphi^I + \sigma^{-1}\omega_2q^I$.  The two contributions recombine nicely and give:
\be\label{eq:AdSBR_action}
\II \,=\, -k_{IJK}q^I\,\frac{3\varphi^J\varphi^K + 3\sigma\,\omega_2\,q^J\varphi^K + \omega_2^2\,q^Jq^K}{6\sigma\omega_1} + k_Iq^I\,\frac{\omega_1^2 +\omega_2^2- 4\pi^2}{24\sigma \omega_1} \,,
\ee
which is just of degree two in $\varphi^I$.
Setting $k_I=0$, this agrees with the two-derivative result of~\cite{Cassani:2025iix,Colombo:2025yqy} for $\sigma = 1$. Beyond that, \eqref{eq:AdSBR_action} provides a prediction for the higher-derivative corrections to 
black ring saddles.

\paragraph{Entropy of asymptotically AdS$_5$ black rings.} In appendix~\ref{app:WaldEntropyRings} we derive the entropy of the related supersymmetric extremal solutions, working at linear order in the higher-derivative corrections.
Since the expressions are rather cumbersome, we report here just the expression for the Bekenstein-Hawking entropy, valid for $\sigma=\pm1$,
\begin{equation}\label{entropy_AdS_BR_bothsigma}
{\cal S} \,=\, \pi\, \sqrt{\frac{ \left(kqqq\right) \left( \big(Q\tilde k^{-1} Q\big) +2\sigma J_1 \right) + 3\left( \left( Qq\right) - 2\sigma J_2\right)^2 }{\big(g\tilde k^{-1}g \big) \left(kqqq\right) + \frac{3}{2}(\sigma+1)}}\,.
\end{equation}
The constraint between the charges is quite different in the two cases: it reads
\begin{equation}
J_2 = \frac{\left(Qq\right)}{2} - \frac{\left(kqqq\right)}{12}\left(1-2\big(g\tilde k^{-1}Q\big)\right)\,,  \qquad  \text{for}\quad \sigma = 1\,,
\end{equation}
\begin{equation}\label{eq:constraint_sigma_minus1}
 \big( g\tilde k^{-1} Q\big) = - \frac{1}{2}\,,\qquad  \text{for}\quad \sigma = -1\,.
\end{equation}
In the formulae above we have introduced the matrix 
\be\label{eq:matrix_tildek}
\tilde k_{IJ} \,=\, k_{IJK} q^K \,,
\ee
that we assume invertible, with inverse $(\tilde k^{-1})^{IJ}$. Moreover,  we use a notation where quantities grouped in parenthesis have their indices contracted in the natural way, namely
\be 
\label{eq:notation_LT}
(Q\tilde k^{-1} Q) = Q_I (\tilde k^{-1})^{IJ} Q_J\,,\qquad\ (k qqq) \,=\, k_{IJK}q^Iq^Jq^K\,,\qquad\ (Qq) = Q_I q^I \,.
\ee

In section~\ref{sec:compare_near_horizon}, we compare our result for $\sigma=-1$ with the Bekenstein-Hawking entropy of the explicit near-horizon geometry constructed in~\cite{Kunduri:2007qy}, finding a remarkable agreement. We will also see that the case $\sigma = -1$ realizes the topological twist on the horizon $S^2$. It follows that the left- and right-moving central charges of the $(0,2)$ CFT$_2$ dual to the AdS$_3$ factor in the near-horizon geometry are given by (at leading order)~\cite{Benini:2013cda}
\be\label{centralcharge_general}
c_L = c_R \,=\,  -\frac{3}{(g\tilde k^{-1} g)}\,,
\ee
which depends on the theory data ($k_{IJK}$, $g_I$) and on the global data of the solution ($q^I$), but not on the conserved charges. The expression for the entropy then takes the form of a charged Cardy formula~\cite{Hosseini:2020vgl},
\begin{equation} 
{\cal S} \,=\, 2\pi\, \sqrt{  \frac{c_L}{6}\, \hat J_1}\,,
\end{equation}
where 
\be
\hat J_1 \,=\,  J_1  -  \frac{1}{2} \big(Q\tilde k^{-1}Q\big) -  \frac{3}{2}  \frac{ \left( 2J_2+\left( Qq\right)\right)^2}{(kqqq)}\, .
\ee
From the equations in appendix~\ref{app:WaldEntropyRings} we also find the extremum value of $\omega_1$ (cf.\ \eqref{eq:expr_omega1_sigma-1}),
\be\label{eq:omega1_maintext}
\omega_1 \,=\, -\frac{\pi^2 c_L}{3\,{\rm Re}\,\mathcal{S}}\,,
\ee
which is valid even without imposing~\eqref{eq:constraint_sigma_minus1}, provided ${\rm Re}\,\mathcal{S}\neq 0$. This is the left-moving inverse temperature of the Cardy formula.

Our conjecture for the higher-derivative corrections to the central charge and to $\hat J_1$ can be read from the expressions for $P_0$ and $P_2$ in appendix~\ref{app:WaldEntropyRings}.

We note that the condition that the 2d central charge be positive constrains the allowed values of the $q^I$. In particular, they cannot be aligned to the R-symmetry. This would mean taking $q^I \propto \bar{X}^I$, with a positive proportionality constant, which would imply $g \tilde k^{-1} g > 0$. 
 It follows that there is no such black ring in minimal gauged supergravity, in agreement with the no-go result of~\cite{Kunduri:2006uh,Grover:2013hja}. This is also consistent with the fact that in the dual large-$N$ field theories, an SCFT$_4 \to$ SCFT$_2$ flow cannot be realized by topologically twisting on $S^2$ unless the 4d R-symmetry mixes with flavour symmetries to realize the 2d R-symmetry~\cite{Benini:2015bwz}.
%

\paragraph{Entropy of asymptotically flat black ring.}  We now focus on the ungauged limit and fix $\sigma = 1$. As mentioned, an asymptotically flat solution with toric symmetry and fan \eqref{eq:fan_BR} can be constructed explicitly at the two-derivative level~\cite{Cassani:2025iix}, serving as saddle for the flat-space gravitational index. In this setup, higher-derivative corrections to the entropy of supersymmetric extremal black rings are known~\cite{Guica:2005ig,Bena:2005ae}. The Wald entropy can be compared with a microscopic evaluation of indicial degeneracies extracted from indices counting Gopakumar-Vafa invariants at large charge, see~\cite{Halder:2023kza,Alexandrov:2026rra} for recent progress. Since in this case a precise comparison is possible, we find it useful to present the details of how the Wald entropy is derived by Legendre transforming our on-shell action.

In the ungauged case we can evaluate the Legendre transform of the black ring action~\eqref{eq:AdSBR_action}  with $\sigma=1$ exactly. The action reads 
\begin{equation}
\II \,=\, -k_{IJK}q^I\,\frac{3\varphi^J\varphi^K + 3\,\omega_2\,q^J\varphi^K + \omega_2^2\,q^Jq^K}{6\omega_1} + k_Iq^I\,\frac{\omega_1^2 +\omega_2^2- 4\pi^2}{24 \omega_1}
\end{equation}
with the constraint $\omega_1+\omega_2 = 2\pi \ii $.\footnote{One may also choose $\omega_1+\omega_2 = -2\pi \ii $. Since the two choices are related by a $4\pi\ii$ shift, they give equivalent boundary conditions, hence the two corresponding saddles should contribute to the same gravitational index.}
 
Let us first compute the action for fixed  electric charges $Q_I$. This is given by
\be
\tilde I(\omega,Q) \,=\, {\rm ext}_\varphi\left[I(\omega,\varphi)  + \varphi^I Q_I \right]\,,
\ee
where ``${\rm ext}_\varphi$'' means that the expression should be evaluated by extremizing with respect to  the electrostatic potentials $\varphi^I$. Substituting the extremum value
\be\label{eq:solext}
\varphi^I \,=\, \omega_1  (\tilde k^{-1})^{IJ}Q_J - \omega_2 \frac{q^I}{2}\,,
\ee
where $\tilde k_{IJ}$ is again the matrix~\eqref{eq:matrix_tildek},
leads to
\be
\tilde I(\omega,Q) \,=\,  \frac{X \omega_1^2 + Y \omega_2^2 +Z \omega_1\omega_2}{\omega_1}\,,
\ee
with 
\begin{align}\label{eq:XYZ}
X\,&=\, \frac{(Q\tilde k^{-1} Q)}{2}  + \frac{(kq)}{12}\,,
\qquad Y\,=\, - \frac{(k qqq) }{24} +  \frac{(kq)}{12}\,, \qquad Z \,=\, -\frac{(Qq)}{2} + \frac{(kq)}{12}\,.
\end{align}
Here we are using the notation \eqref{eq:notation_LT}, and $(kq)=k_Iq^I$.

The entropy is now given by the constrained Legendre transform
\be
{\cal S}(J,Q) \,=\, {\rm ext}_{\{\omega_1,\omega_2,\Lambda\}} \left[  - \tilde I(\omega,Q) - \omega_1J_1 -\omega_2J_2 - \Lambda (\omega_1+\omega_2 -2\pi i)  \right]\,.
\ee
It is easy to see that
$ 
{\cal S} = 2\pi i \Lambda\,,
$ 
where 
 $\Lambda$ is determined by the quadratic equation
\be\label{eqforLambda_ring}
P_0 + P_1  \Lambda + \Lambda^2 \,=\, 0\,,
\ee
with
\be\label{P0P1_ring}
P_0 \,=\, (Z + J_2)^2 - 4Y (X+J_1)\,,\qquad\quad P_1 \,=\, 2 (J_2 - 2Y +Z)\,.
\ee
Solving the equation by demanding in addition that $\cal S$ is real and positive leads to
\be
{\cal S} \,=\, 4\pi \sqrt{-Y (X-Y+J_1)}\,,
\ee
with 
\be
J_2 \,=\, 2Y-Z\,.
\ee
Plugging the values \eqref{eq:XYZ} in, one arrives at
\be
\label{eq:AFBR_entropy}
{\cal S} \,=\, 2\pi \sqrt{ \frac{(k qqq)  -  2(kq)}{6} \left[J_1  + \frac{(k qqq) }{24}  +  \frac{1}{2}(Q\tilde k^{-1} Q)  \right]}\,,
\ee
with 
\be
\label{eq:AFBR_charges}
J_2 \,=\,  \frac{(Qq)}{2}  - \frac{(k qqq) -(kq)}{12}   \,.
\ee

This expression for the entropy is in agreement with~\cite{Guica:2005ig,Bena:2005ae}, including the higher-derivative corrections.\footnote{
Our expression for the entropy matches eqs.~(5)--(9) of~\cite{Guica:2005ig} by taking
$$
q^I = p^A|_{\rm there} \,,\qquad Q_I = q_A |_{\rm there}\,,\qquad  k_{IJK} = 6 D_{ABC}|_{\rm there}\,,\qquad J_1 = -J_\psi|_{\rm there}\,,\qquad k_I = - \frac{1}{2}c_{2A}|_{\rm there}\,.
$$
One also finds $\left(kqqq\right) \,=\, 6D_{\rm there}$ and $(Q\tilde k^{-1} Q)  + \frac{(k qqq) }{12} +2J_1 \,=\, 2\,\hat q_0|_{\rm there}$.
}
The expression for $J_2$ agrees with~\cite{Bena:2005ae} up to the higher-derivative $(kq)$ correction, which does not seem to have appeared before.\footnote{In particular, it does not appear in the AdS$_3\times S^2$ near-horizon analysis of~\cite[sect.~8]{deWit:2009de}, which extends to higher-derivative couplings the approach of~\cite{Hanaki:2007mb} for evaluating the charges.}
Moreover, by using \eqref{eq:AFBR_charges}, we may rewrite \eqref{eq:AFBR_entropy} as 
\begin{equation}
{\cal S} = 2\pi \sqrt{\frac{\left(kqqq\right) - 2\left(kq\right)}{6}\left[ J_1 - J_2 - \frac{\left(kqqq\right)-2\left(kq\right)}{24} + \frac{\left(Qq\right)}{2} + \frac{1}{2}(Q\tilde k^{-1}Q)\right]}\,,
\end{equation}
which makes the connection with the recent results of~\cite{Alexandrov:2026rra} more manifest.\footnote{This rewriting of the entropy matches eq.~(3.13) of~\cite{Alexandrov:2026rra} upon using the following map:
$$
J_1-J_2 = n_{\rm there}\,,\quad\ \left(kqqq\right) = \kappa r^3|_{\rm there}\,,\quad\ \left(kq\right) = -\frac{1}{2}c_2 r|_{\rm there}\,,\quad\ \left(Qq\right) = rd|_{\rm there}\,,\quad\ (Q\tilde k^{-1}Q) = \frac{d^2}{\kappa r}\Big|_{\rm there}\,.
$$
In particular, one finds $J_1 - J_2 - \frac{\left(kqqq\right)-2\left(kq\right)}{24} + \frac{\left(Qq\right)}{2} + \frac{1}{2}(Q\tilde k^{-1}Q) \,=\, -\hat q_0|_{\rm there}$.
}   
Note that $c_L = (k qqq)  -  2(kq)$ and $c_R = (k qqq)  -  (kq)$ are the left- and right-moving central charges of the $(0,4)$ CFT$_2$ living at the boundary of the AdS$_3$ in the near-horizon geometry of the asymptotically flat BPS black ring, while  $ -\tilde k_{IJ}$ is the level matrix of the flavour charges.

Finally, we note that one can solve~\eqref{eqforLambda_ring} in a more trivial way, by setting $P_0=\Lambda=0$, which leads to a vanishing entropy, ${\cal S}=0$.  As shown in~\cite{Cassani:2025iix}, at the two-derivative level this gives a topological soliton, also known as three-center horizonless geometry~\cite{Bena:2005va,Bena:2007kg}. The condition $P_0 =0$, with $P_0$ given by \eqref{P0P1_ring}, should thus correspond to the relation between the charges satisfied by the topological soliton in the presence of higher-derivative corrections.


\subsection{Black hole in bubbling spacetime}

As last example, we consider a geometry with \(s=4\), namely with four nuts. We take the fan to be given by
\begin{equation}
\label{eq:fan_BHBS}
V^0=\left(0,1,0\right)\,,
\quad
V^1=\left(1,0,0\right)\,,
\quad
V^2=\left(0,0,1\right)\,,
\quad
V^3=\left(0,1,0\right)\,,
\quad
V^4=\left(0,0,1\right)\,.
\end{equation}
The compact fixed loci correspond to a horizon, ${\cal D}_1\cong S^3$, together with two bubbles, ${\cal D}_2\cong S^3$ and ${\cal D}_3\cong S^1\times S^2$. In the ungauged limit, this geometry therefore provides a supersymmetric non-extremal version of the BPS black hole in a bubbling spacetime constructed explicitly in~\cite{Horowitz:2017fyg}. In this section we compute the on-shell action of the candidate asymptotically AdS$_5$ geometry with the same topology. Taking the ungauged limit then gives a prediction for the higher-derivative correction to the action of the solution of~\cite{Horowitz:2017fyg}. 
%
%
%

The regularity constraints \eqref{eq:int_constraint} impose
\begin{equation}
q^{\cal R}_2\,=\,0\,=\, q^{\cal R}_3\,,
\end{equation}
while we are also taking $q_1^I = 0$. Therefore, the fluxes that characterize the Chern classes \eqref{eq:gauge_chern} are given by
\begin{equation}
\DD_0^I = 0 = \DD_4^I \,,\qquad \DD_1^I = \frac{\varphi^I}{2\pi\ii}\,,\qquad \DD_2^I = -q_2^I\,,\qquad \DD_3^I = - q_3^I\,.
\end{equation}

\paragraph{Six-dimensional extension and on-shell action.}

The fan \eqref{eq:fan_BHBS} can be represented by a degenerate diagram with points
\begin{equation}
v^0=\left(0,0\right)\,,
\qquad
v^1=\left(0,1\right)\,,
\qquad
v^2=\left(1,0\right)\,,
\qquad
v^3=v^0\,,
\qquad
v^4=v^2\,.
\end{equation}
A triangulation of this diagram gives the three cones\footnote{As in the black ring case, the toric fan of the black hole in bubbling spacetime is degenerate. We remove this degeneracy by introducing a deformation of the fan, such that it describes a convex pentagon with vertices $\left(0,0\right)$, $\left(0,1\right)$, $\left(p_2,n_2\right)$, $\left(p_3,n_3\right)$ and $\left(1,0\right)$ (ordered clockwise). Among the possible triangulations of this pentagon, there is one whose associated cones remain non-singular in the limit that recovers the original fan, i.e. $p_2\to 1$, $n_2 \to 0$ and $p_3 \to 0$, $n_3\to 0$.}
\begin{equation}
\CC_1=\{V^1,V^2,V^3\}\,,
\qquad
\CC_2=\{V^0,V^1,V^4\}\,,
\qquad
\CC_3=\{V^1,V^3,V^4\}\,.
\end{equation}
We quote here the final result for the integral of the anomaly polynomial:
\begin{equation}
\label{eq:action_BHBS}
\begin{aligned}
\II\,&=\, \frac{k_{IJK}}{6} \Bigl[\frac{\varphi^I\varphi^J\varphi^K}{\omega_1\omega_2} +\frac{q_2^I}{\omega_1}\Bigl( 3\left(\varphi^J + \omega_1 q_3^J\right) \left(\varphi^K + \omega_1 q_3^K\right) + 3 \omega_2 q_2^J\left(\varphi^K + \omega_1 q_3^K\right) + \omega_2^2 q_2^J q_2^K \Bigr)
\Bigr]
\\[1mm]
&\quad\, - k_I\left(\varphi^I + \omega_2q_2^I\right)\frac{\omega_1^2+\omega_2^2-4\pi^2}{24\omega_1\omega_2}\,.
\end{aligned}
\end{equation}

In appendix \ref{app:BHinBS} we calculate the Wald entropy and charge constraint of the related asymptotically flat  supersymmetric and extremal solution. In particular, we verify that the leading order expressions agree with the Bekenstein-Hawking entropy and charge constraint originally derived in~\cite{Horowitz:2017fyg}.

\section{AdS$_5$ black ring entropy and near-horizon solution}\label{sec:compare_near_horizon}

In this section, we consider the ${\rm AdS}_3\times S^2$ solution to two-derivative gauged supergravity 
 given in~\cite{Kunduri:2007qy}, which may be seen as the near-horizon geometry of a black ring in AdS$_5$. We determine the conserved charges of the solution and derive an expression for the Bekenstein-Hawking entropy as a function of the charges. We will find that this agrees precisely with the formula for the entropy that was obtained in section~\ref{sec:black_rings} by Legendre-transforming the AdS$_5$ black-ring on-shell action with $\sigma = -1$. We will also clarify how the choice $\sigma = -1$ is related to the topological twist on $S^2$.

Our context is the five-dimensional R-gauged supergravity of section~\ref{sec:sugra_sol}, with action \eqref{eq:R-gauged_action}.
Here we work in Lorentzian signature and use the supergravity tensor $C_{IJK}$, converting to $k_{IJK} $ via \eqref{eq:Ctok} only at the end.
 For concreteness, we shall refer to the theory with three gauge fields, $I=1,2,3$, that arises as a consistent truncation of type IIB supergravity on~$S^5$.
This is obtained by taking $C_{IJK} = |\epsilon_{IJK}|$ and fixing the Fayet-Iliopoulos parameters  as $g_I = \frac{1}{2}g$.
Then the scalars $X^I$ satisfy the constraint $X^1 X^2 X^3 =1$, the kinetic matrix is
\be
Q_{IJ} = \frac12 \,{\rm diag} \left[ (X^1)^{-2} , (X^2)^{-2} , (X^3)^{-2}  \right],
\ee
 the  scalar potential reads
\begin{equation}
{\cal V} = - 2 g^2 \left[  (X^1)^{-1}  +  (X^2)^{-1}  +  (X^3)^{-1} \right],
\end{equation}
and is 
extremized at the scalar value $\bar X^I = 1$, with  $g^{-1}$ the AdS$_5$ radius. Our discussion below relies on this specific theory just for a few expressions; the generalization to the case with an arbitrary number of vector multiplets and a symmetric scalar manifold is straightforward.
%
%

\paragraph{The near-horizon solution.} The theory admits a smooth AdS$_3\times S^2$ solution, which may be interpreted as the near-horizon geometry of a black ring~\cite{Kunduri:2007qy}.  
 The metric reads
\begin{equation}\label{nearhorizon_metric}
\diff s^2 = 2 \diff v \diff r - 2 g X \frac{L}{2\pi}\, r\,\diff v \diff \psi + \frac{L^2}{4\pi^2}\diff \psi^2 + \frac{1}{g^2\lambda}\left( \diff \theta^2 + \sin^2\theta \diff \phi^2\right)\,,
\end{equation}
where $\theta \in [0,\pi]$, $0\leq \phi <2\pi$ parameterize $S^2$, and $0\leq \psi < 2\pi$ is a coordinate on an $S^1$ with proper length $L$. 
This describes the direct product of the near-horizon limit of an extremal BTZ black hole and a two-sphere.
 The horizon is located at $r=0$, the spatial sections have $S^1 \times S^2$ topology and the Bekenstein-Hawking entropy reads
\begin{equation}
\label{eq:BRBHE}
{\cal S} \,=\,\frac{{\cal A}_{\cal H}}{4G} \,=\,  \frac{\pi L}{G g^2\lambda}\,.
\end{equation}
 Here, the scalars are constant and we have introduced the combinations
\begin{equation}
X \,=\, X^1 + X^2 + X^3\,,
\end{equation}
\begin{equation}
\lambda = \left(X^1\right)^2 + \left(X^2\right)^2 + \left(X^3\right)^2 - 2 X^1 X^2 - 2 X^1 X^3 - 2 X^2 X^3\,.
\end{equation}

The gauge field strengths read\footnote{The parameters $q^I$ appearing here are related to those in~\cite{Kunduri:2007qy} as $q^I=- 2q^I_{\rm there}$. We have fixed the notation so that these $q^I$ can be identified precisely with those used in the previous sections, as it will be proven below.} 
\be
F^I \,=\, -\frac{q^I}{2}\,\sin\theta\,\diff\theta\wedge \diff \phi\,,
\ee
hence there is non-trivial flux through the horizon $S^2$,
\begin{equation}\label{defqI_KL}
q^I = -\frac{1}{2\pi}\int_{S^2}F^I\,.
\end{equation}
The $q^I$ provide the dipole charges of the black ring. 
They are related to the scalars as
\begin{equation}\label{eq:qI_from_XI}
q^I \,=\, -\frac{2}{g\lambda}\,X^I\left( X - 2 X^I\right) \,,
\end{equation}
where there is no sum over the repeated indices. Importantly, they satisfy the relation
\begin{equation}\label{eq:constr_dipole}
g_I q^I \,=\, 1\,,
\end{equation}
specifying that the R-flux through $S^2$ must be $-1$, thus implementing the topological twist. 

Because of the non-trivial flux, the gauge potential needs to be defined patchwise. We will use the same patches identified in~\cite{Hanaki:2007mb} for the near-horizon geometry of the asymptotically flat black ring. Following the notation therein, we introduce a patch $S$ spanning $-1 \leq \cos\theta < 1-\epsilon$ (containing the south pole of $S^2$) and a patch $T$ for $1-\epsilon < \cos\theta \leq 1$ (containing the north pole of $S^2$), with $\epsilon$ a small parameter that will be sent to zero in the computations below. We define the gauge fields in each patch as
\begin{equation}\label{gauge_patches_BR}
\begin{aligned}
A_S^I \,&=\,   \frac{q^I}{2} \left( \cos\theta +1 \right) \diff \phi + \left( a^I - \frac{b^I}{2}\right) \diff \psi\,,\\[1mm]
A_T^I \,&=\, \frac{q^I}{2} \left( \cos\theta -1 \right) \diff \phi + \left( a^I + \frac{b^I}{2}\right) \diff \psi\,,
\end{aligned}
\end{equation}
hence the gauge transformation on the patch interface reads
\be\label{eq:gauge_transfST}
\zeta^I \,\equiv\, A_S^I -A_T^I  \,=\, q^I \diff\phi - b^I \diff\psi \,.
\ee 
In~\eqref{gauge_patches_BR}, we have introduced a constant $\diff\psi$ component, and parameterized its possible different values in the $S$ and $T$ patches by the constants $a^I$ and $b^I$. These parameters are arbitrary in the near-horizon geometry and were not considered in~\cite{Kunduri:2007qy}.

Before moving on, we note that $\lambda$ must be positive for the $S^2$ area to be positive. It follows that one cannot take the limit to minimal gauged supergravity, since this requires $X^1=X^2=X^3=1$, which implies $\lambda = -3$~\cite{Kunduri:2007qy}.
The Brown-Henneaux formula gives the following expression for the left- and right-moving central charges of the dual CFT$_2$:
\be\label{eq:c_AdS_ring}
c_L \,=\, c_R\,=\, \frac{3\ell_3}{2 G_3} \,=\, \frac{12\pi}{ G g^3 \lambda X}    \,,
\ee
where $\ell_3 =\frac{2}{gX} $ is the AdS$_3$ radius and $G_3 = \frac{G}{ {\rm Vol}(S^2)} = \frac{Gg^2 \lambda}{4\pi }$ is the three-dimensional Newton's constant.
So positivity of $\lambda$ can also be seen as positivity of the central charge, reproducing the condition found analyzing the Legendre transform of the on-shell action.\footnote{We also note that $c_L = \frac{96}{\lambda X } a_{\rm 4d}$, where $a_{\rm 4d}=\frac{\pi}{8Gg^3}$ is the central charge of the SCFT$_4$ dual to ${\rm AdS}_5$, which for type IIB on $S^5$ is $\mathcal{N}=4$ SYM.}
Indeed, it is not hard to see that
\be\label{eq:gCtildeg}
(g\tilde C^{-1} g)=  -\frac{g^3 \lambda X}{16}\,,
\ee
where we introduced the matrix 
\begin{equation}\label{eq:tildeC}
\tilde C_{IJ}\, = \,  C_{IJK} q^K\,,
\end{equation}
which we assume invertible. Eq.~\eqref{eq:gCtildeg} confirms that  the central charge~\eqref{eq:c_AdS_ring} extracted from the near-horizon solution agrees with the general expression~\eqref{centralcharge_general}.

\paragraph{Constraints from the full geometry.} Although the full asymptotically AdS$_5$ solution is not known, we can use the  toric data specified in section~\ref{sec:black_rings} to constrain the free parameters in the near-horizon solution. 
In order to do so, we need to relate the angular coordinates $\phi,\psi$ above to the $2\pi$-periodic angles $\phi_1$, $\phi_2$ that we use to define the fan vectors and therefore the global geometry. We take
\be\label{eq:change_coords_nh}
\phi_1 = \psi \,, \qquad \phi_2 = \phi - c \,\psi\,,
\ee
where $c$ is a constant. The identifications of $\phi_1,\phi_2$ imply the following identifications of the near-horizon coordinates,
\be
(\phi, \psi) \,\sim\, ( \phi + 2\pi, \psi ) \,\sim\, ( \phi +2 \pi c, \psi + 2\pi )\,,
\ee
For non-integer $c$, these are twisted identifications, namely going around $S^1$ also involves a rotation in $S^2$.\footnote{Equivalently, we could have worked with the untwisted coordinates $\phi_1,\phi_2$ and introduced a constant connection proportional to $c$ in the metric \eqref{nearhorizon_metric}, fibering the circle in $S^2$ over $S^1$.} 
In terms of vectors, we have the map:
\be\label{dictionary_NH_vectors}
\partial_\phi = \partial_{\phi_2}\,,\qquad \partial_\psi = \partial_{\phi_1} - c \,  \partial_{\phi_2}\,.
\ee

We also need to make some assumptions on the gauge patches in the full geometry. We assume that the $S$ and $T$ patches introduced in the near-horizon geometry of the extremal solution can be extended to the full geometry of the non-extremal supersymmetric solution, in such a way that the $T$ patch incorporates the compact bolt $\mathcal{D}_1$  and slightly stretches both into the horizon $\mathcal{D}_2$ and into $\mathcal{D}_0$, while the $S$ patch is the complement of the closure of the $T$ patch, so it covers most of 
$\mathcal{D}_2$ and  of the semi-infinite bolt $\mathcal{D}_0$, as well as the whole semi-infinite bolt $\mathcal{D}_3$, and reaches the conformal boundary. Since both $S$ and $T$ intersect the bolt $\mathcal{D}_0$, 
 so does the patch interface.

With these assumptions, we can see that the definition~\eqref{defqI_KL} of the $q^I$ as $S^2$-fluxes agrees with the definition of $q^I$ as flat connections used in the previous sections, cf.\ eq.~\eqref{eq:qaI}. For the black ring, the latter gives $q^I = - \left. V^1\cdot A^I_S \right |_{{\cal D}_1}$, where the gauge field is taken in the $S$ patch since by assumption this is the one extending to the conformal boundary. 
This expression evaluates to
\be\label{eq:qisminushatq}
- \left. V^1\cdot A^I_S \right |_{{\cal D}_1} \,=\,  \left.(A^I_{S})_{\phi_2} \right |_{\theta=0} \,=\,  \left.(A^I_{S})_{\phi} \right |_{\theta=0} \,,
\ee
 where we used $V^1 = \sigma \partial_{\phi_2} = -  \partial_{\phi_2}$, and we noted that evaluating the gauge field component  at ${\cal D}_1$ is equivalent to evaluate it at $\theta=0$ in the near-horizon geometry. Comparing with the explicit form of the near-horizon gauge field in \eqref{gauge_patches_BR}, we see that the component on the right hand side is precisely the  $q^I$ measuring (minus) the $S^2$ flux.
 This establishes the identity between the two definitions of $q^I$.
It follows that the constraint \eqref{eq:constr_dipole} on the dipole charges is the same as the one obtained from the global analysis in section~\ref{sec:black_rings}.

We now illustrate how the constants introduced in the near-horizon gauge field are constrained.
Since we are assuming that the interface between the $S$ and $T$ patches intersects $\mathcal{D}_0$, and  $V^0=\partial_{\phi_1}$, regularity of the gauge field at $\mathcal{D}_0$ requires $\partial_{\phi_1} \cdot \zeta^I =0$, which determines
\be\label{eq:expr_bI}
b^I \,=\,  c\, q^I\,.
\ee
Note that the gauge transformation~\eqref{eq:gauge_transfST} can then be expressed as $\zeta = q^I \diff\phi_2$.
Next we explain how a combination of the $a^I$, namely the R-symmetry combination $g_I a^I$,  is fixed. The argument is based on regularity of the Killing spinor, that we denote by $\varepsilon$. 
When taking the near-horizon extremal limit, the Killing spinor of the full solution should reduce to the one of the near-horizon limit of extremal  ${\rm BTZ}\times S^2$. In the latter geometry, besides realizing the topological twist on $S^2$, the spinor must be independent of the coordinate $\psi$ parameterizing the $S^1$ in a gauge where the R-gauge field has vanishing $\psi$-component, namely it must agree with the periodic Killing spinor of the extremal BTZ factor~\cite{Coussaert:1993jp,Coussaert:1994tu}. The gauge-invariant version of this statement uses the gauge-covariant spinorial Lie derivative and reads
\be\label{gauge_cov_Lie}
\left(\mathcal{L}_{\partial_\psi} - \ii\, \partial_\psi \cdot (g_I A^I )\right) \varepsilon \,=\, 0\,,
\ee
where $g_I A^I$ is the R-symmetry gauge field and the expression holds in the near-horizon solution.\footnote{Equivalently, one can consider the one-form $W$ in the analysis of~\cite[sec.~3.3.1]{Kunduri:2007qy}, which originates from a spinor bilinear and has R-charge $-2$. In the solution under study, eq.~(51) of~\cite{Kunduri:2007qy} reads $\diff W + 2\ii \,g_I A^I \wedge W=0$. This equation implies that $W$, and thus the Killing spinor, does not depend on $\psi$ in a gauge where $g_I A^I_\psi=0$. It also encodes the topological twist on $S^2$.}
Let us evaluate \eqref{gauge_cov_Lie} in the patch $S$. In this patch, the gauge field is regular at the fixed loci of $V_0 = \partial_{\phi_1}$  and  $V_3 = \partial_{\phi_2}$ and vanishes at infinity, hence a regular spinor must be antiperiodic along the contractible circles defined by the fan vectors. Namely, it satisfies
\be
\mathcal{L}_{\partial_{\phi_1}}\varepsilon_S \,=\, \frac{\ii}{2} \, \varepsilon_S\,, \qquad \mathcal{L}_{\partial_{\phi_2}}\varepsilon_S \,=\, \frac{\ii}{2}\,  \varepsilon_S\,.
\ee
  Here the signs are fixed by the choice
~\eqref{eq:int_constraint}, which implies that the Killing spinor has the same eigenvalue $+\frac{\ii}{2}$ under the action of any fan vector at the corresponding bolt in the gauge that is regular there. Then, recalling  \eqref{dictionary_NH_vectors} we have,
\be\label{spinorial_der_S}
\mathcal{L}_{\partial_\psi}\varepsilon_S \,=\, \frac{\ii}{2}\left(1-c\right) \varepsilon_S
\ee
and, from \eqref{gauge_patches_BR},
\be\label{contration_A_S}
\partial_\psi \cdot (g_IA^I_S)\,=\,   g_I \left(a^I -  \frac{c q^I}{2}\right) \,= \,  g_Ia^I  - \frac{c}{2}\,. 
\ee
Hence condition \eqref{gauge_cov_Lie} requires
\be\label{eq:constraint_a}
g_I a^I\, =\, \frac{1}{2}\,. 
\ee 
Note that the factors of $c$ cancel out in the gauge-covariant Lie derivative. Although covariance guarantees that the same result is obtained in the $T$ patch, it is an instructive consistency check to see how this works. The gauge transformation due to the R-flux implies that $c$ appears with an opposite sign in the $T$-patch versions of~\eqref{spinorial_der_S} and~\eqref{contration_A_S}. It also implies that in this patch
\be
\mathcal{L}_{\partial_{\phi_1}}\varepsilon_T \,=\, \frac{\ii}{2} \, \varepsilon_T\,, \quad\text{but}\quad \mathcal{L}_{\partial_{\phi_2}}\varepsilon_T \,=\, -\frac{\ii}{2}\,  \varepsilon_T\,.
\ee
Rewriting the second as  $\mathcal{L}_{-\partial_{\phi_2}}\varepsilon_T \,=\, \frac{\ii}{2}\,\varepsilon_T$ expresses the fact that the fan vector of the compact bolt $\mathcal{D}_1$ (which is in the $T$-patch) is $V^1 = - \partial_{\phi_2}$, consistently with the sign choice $\sigma=-1$.

In summary, the parameters $a^I$ and $b^I$ introduced in~\eqref{gauge_patches_BR}, which are free in the near-horizon geometry, are constrained by the AdS$_5$ embedding as in~\eqref{eq:expr_bI} and~\eqref{eq:constraint_a}.  Overall, the solution depends on the $S^1$ length $L$,  the twist parameter $c$, the dipole charges $q^I$ and the holonomies $a^I$, $I=1,2,3$, subject to the constraints~\eqref{eq:constr_dipole} and~\eqref{eq:constraint_a}, which makes six free parameters in total.

Note that  the parameter counting is different in the explicitly known asymptotically flat black ring solution to the ungauged theory \cite{Elvang:2004rt,Elvang:2004ds}. Since the Killing spinor is uncharged, the $S^1$-periodicity condition simply reads $\mathcal{L}_{\partial_\psi}\varepsilon=0$, which is solved by $c=1$, a condition that is indeed satisfied in that solution (this can be seen from \cite{Hanaki:2007mb}, after converting to our conventions). On the other hand, the constraints~\eqref{eq:constr_dipole} and~\eqref{eq:constraint_a} do not arise, hence there are seven free parameters in that case.

\paragraph{Page electric charges.} Our goal is to  express the entropy as a function of the conserved charges carried by the solution --- the electric charges $Q_I$ and the angular momenta $J_{1,2}$ --- and of the dipole charges $q^I$. 
In general, conserved charges are determined by asymptotic integrals in the full solution. However, by expressing them as suitable integrals of closed three-forms, they can be evaluated at the horizon rather than at infinity.
This is the method of~\cite{Hanaki:2007mb}.
 For instance, for the electric charge this goes as follows:
\begin{equation}
\begin{aligned}\label{eq:electric_page}
Q_I &= \frac{1}{8\pi G}\int_{\Sigma_{\infty}} Q_{IJ} \star F^J = \frac{1}{8\pi G}\int_{\Sigma_\infty} \left[Q_{IJ} \star F^J + \frac{1}{4}C_{IJK} A^J \wedge F^K\right] 
\\[1mm]
&= \frac{1}{8\pi G}\int_{\Sigma_r} \left[Q_{IJ} \star F^J + \frac{1}{4}C_{IJK} A^J \wedge F^K\right]\,,
\end{aligned}
\end{equation}
where $\Sigma_\infty \subset \partial {\cal M}$ is a three-dimensional spatial hypersurface in the conformal boundary (we assume the conformal boundary is located at $r = \infty$, where $r$ is the radial coordinate), while  $\Sigma_r$ is a finite-$r$ three-dimensional hypersurface such that $\Sigma_\infty =\lim_{r\to\infty} \Sigma_r$, and such that there exists a four-dimensional chain $\hat\Sigma$ satisfying $\partial \hat \Sigma = \Sigma_\infty - \Sigma_r$.
Eq.~\eqref{eq:electric_page} starts with the usual definition of the Maxwell electric charge. In the second step, we used that for asymptotically AdS$_5$ solutions the combination $A^I \wedge F^K$ is suppressed asymptotically and does not contribute, while in the third step we used Stokes' theorem and Maxwell's equation to convert the integral over $\Sigma_\infty$ to an integral over $\Sigma_r$. The final expression is the definition of the Page electric charge, which in this case agrees with the Maxwell charge. In particular, it can be evaluated at the horizon hypersurface $\Sigma_0$. In evaluating the integrals over $\Sigma_0$, we will use the orientation $ \diff\theta\wedge \diff\phi \wedge \diff \psi$.
 
The Page charge involves a Chern-Simons term, hence care is needed whenever the gauge field is not globally defined, as in the case of the black ring. 
 As argued in~\cite{Hanaki:2007mb}, the evaluation of the Chern-Simons term involves the sum of the contributions over the two patches plus a compensating term localized on the interface between them,
\begin{equation}
\int_{\Sigma_0} C_{IJK} A^J \wedge F^K = \int_{\Sigma_0\cap S} C_{IJK} A_S^J \wedge F^K + \int_{\Sigma_0\cap T} C_{IJK} A_T^J \wedge F^K + \int_{\partial S \cap \Sigma_0} C_{IJK} A_S^J \wedge \zeta^K\,
\end{equation}
where the orientation on $\partial S \cap \Sigma_0$ is $\diff \phi\wedge \diff \psi$.
With this prescription, the electric charge is given by
\begin{equation}
\begin{aligned}
Q_I &= \frac{1}{8\pi G}\int_{\Sigma_0} \left[Q_{IJ} \star F^J + \frac{1}{4}C_{IJK} A^J \wedge F^K\right]
\\[1mm]
&= \frac{1}{8\pi G}\int_{\Sigma_0} Q_{IJ} \star F^J + \frac{1}{32\pi G}\left[\int_{\Sigma_0\cap S} C_{IJK} A_S^J \wedge F^K + \int_{\Sigma_0\cap T} C_{IJK} A_T^J \wedge F^K\right]
\\[1mm]
&\qquad  + \frac{1}{32\pi G}\int_{\partial S \cap \Sigma_0} C_{IJK} A_S^J \wedge \zeta^K\,,
\end{aligned}
\end{equation}
that we will evaluate in the $\epsilon\to 0$ limit.
 The first term pulls back to zero on $\Sigma_0$, while the third term vanishes in the limit, 
hence we are left with~\cite{Hanaki:2007mb}
\begin{equation}
Q_I = \frac{1}{32\pi G} \lim_{\epsilon \to 0 } \left[\int_{\Sigma_0\cap S} C_{IJK} A_S^J \wedge F^K + \int_{\partial S \cap \Sigma_0} C_{IJK} A_S^J \wedge \zeta^K\right]\,.
\end{equation}
This evaluates to
\begin{equation}\label{eq:rel_Qa}
\begin{aligned}
Q_I \,&=\, -\frac{\pi}{16G}\left[C_{IJK} q^J\left(2a^K - c q^K\right) + C_{IJK} q^J\left( 2a^K + c q^K\right) \right]\\[1mm]
 \,&=\, -\frac{\pi}{4G}\tilde C_{IJ} a^K\,.
\end{aligned}
\end{equation}
Therefore, the electric charges of the solution are related to the $a^I$ holonomies in the near-horizon gauge field through the matrix $\tilde C_{IJ}$ defined in~\eqref{eq:tildeC}. Then the condition \eqref{eq:constraint_a} implies the following constraint between the electric charges:
\begin{equation}\label{eq:constraint_on_charges_0}
\frac{4G}{\pi}\,g_I (\tilde C^{-1})^{IJ} Q_J \,=\, - \frac12 \,.
\end{equation}

\paragraph{Angular momenta.} A similar discussion applies to the angular momenta. In this case, one finds the integrals~\cite{Hanaki:2007mb,Cassani:2018mlh} 
\begin{equation}
J_{[\xi]} \,=\, -\frac{1}{16\pi G}\int_{\Sigma_r} \left[ \star\, \diff \xi^\flat + 2\, \iota_\xi A^I \left( Q_{IJ} \star F^J + \frac{1}{6}C_{IJK} A^J \wedge F^K\right) \right]\,,
\end{equation}
where $\xi$ is a Killing vector and $\xi^\flat \equiv \xi_\mu \diff x^\mu$. At infinity, the part depending on the gauge field does not contribute and the integral reduces to the first term, that is the usual Komar integral defining the angular momentum of the full spacetime. Again, we will evaluate the integral at the horizon, $\Sigma_0$. As above, this integral requires considering the sum of the contributions from the two patches plus a compensating term from the patch interface. Following~\cite{Hanaki:2007mb}, this gives
\begin{equation}
\begin{aligned}
J_{[\xi]} &= -\frac{1}{16\pi G}\int_{\Sigma_0} \star\, \diff \xi^\flat - \frac{1}{8\pi G}\int_{\Sigma_0\cap S} \iota_\xi A_S^I \left(Q_{IJ} \star F^J + \frac{1}{6}C_{IJK} A_S^J \wedge F^K\right)
\\[1mm]
&\quad -  \frac{1}{8\pi G}\int_{\Sigma_0\cap T} \iota_\xi A_T^I \left(Q_{IJ} \star F^J + \frac{1}{6}C_{IJK} A_T^J \wedge F^K\right)
\\[1mm]
&\quad - \frac{1}{8\pi G}\int_{\Sigma_0\cap \partial S}\left[ \iota_\xi \zeta^I\,Z_I - \frac{1}{6}C_{IJK}\left( \iota_\xi \zeta^I + \frac{1}{2}\iota_\xi A_T^I \right) \zeta^J \wedge A^K_T\right]
 \,,
\end{aligned}
\end{equation}
where $Z_I$ satisfies
\begin{equation}
\diff Z_I = Q_{IJ} \star F^J +\frac{1}{4} C_{IJK} A_T^J \wedge F^K\,.
\end{equation}
Again the term involving $\star F^I$ does not contribute. Moreover, in the $\epsilon \to 0$ limit the second line and the term involving $Z_I$ in the third line vanish and can be discarded.

We first compute $J_{[\phi]}$. It is straightforward to show that the term comprising $\star\, \diff (\partial_\phi)^\flat$ vanishes, while also $\lim_{\epsilon \to 0}\iota_{\partial_{\phi}} A_T^I=0$. Then, the expression reduces to
\begin{equation}
J_{[\phi]} \,=\, \frac{C_{IJK}}{48\pi G}\left[- \int_{\Sigma_0 \cap S} (\iota_{\partial_{\phi}} A_S^I) \,A_S^J \wedge F^K + \int_{\Sigma_0 \cap \partial S} (\iota_{\partial_{\phi}} \zeta^I) \,\zeta^J \wedge A_T^K \right]\,,
\end{equation}
which evaluates to
\begin{equation}
\begin{aligned}
J_{[\phi]} \,&=\,\frac{\pi}{48 G}\left[ C_{IJK} q^I q^J \left( 2a^K - c q^K\right) + 2C_{IJK} q^I q^J \left( 2a^K + c q^K\right)\right] 
\\[1mm]
&= \frac{\pi}{48G}\left[ 6 (C a qq) + c \,(Cqqq)\right]\,,
\end{aligned}
\end{equation}
where we use the notation
$(Cxyz) \,=\, C_{IJK} x^I y^J z^K\,.$
The formula for the other conserved angular momentum reads
\begin{equation}
\begin{aligned}
J_{[\psi]} \,&=\, - \frac{1}{16\pi G}\int\star\,\diff \left(\partial_\psi\right)^\flat - \frac{C_{IJK}}{48\pi G}\int_{\Sigma_0 \cap S} (\iota_{\partial_{\psi}} A_S^I )\,A_S^J \wedge F^K
\\[1mm]
\,&\, \quad\, +  \frac{C_{IJK}}{48\pi G}\int_{\Sigma_0 \cap \partial S} \left( \iota_{\partial \psi} \zeta^I + \frac{1}{2}\iota_{\partial\psi} A_T^I\right) \zeta^J \wedge A_T^K\,,
\end{aligned}
\end{equation}
which, using 
\be
\int_{\Sigma_0}\star\,\diff \left(\partial_\psi\right)^\flat = - \frac{ 2 L^2 X}{g\lambda}\,,
\ee 
evaluates to
\begin{equation}
J_{[\psi]} \,=\,  \frac{L^2 X}{8\pi G g\lambda} + \frac{\pi}{8G} (C a a q) - c J_{[\phi]} + \frac{\pi c^2}{96G}\,(Cqqq)\,.
\end{equation}
In order to compare with our prediction for the black ring entropy we need the angular momenta associated with $\partial_{\phi_1}$ and $\partial_{\phi_2}$. These are given by
\begin{equation}\label{eq:exprJ1J2_NH}
\begin{aligned}
J_2 \,&=\,  J_{[{\phi}]} \,=\, -\frac{1}{2}(Qq)  + \frac{\pi c }{48G}\,(Cqqq)\,,
\\[1mm]
J_1 \,&=\, J_{[\psi]} + c J_{[\phi]} \,=\, \frac{L^2 X}{8\pi G g\lambda}+ \frac{2G}{\pi }(Q\tilde C^{-1}Q) + \frac{\pi c^2}{96G}\,(Cqqq)\,,
\end{aligned}
\end{equation}
where we used \eqref{eq:rel_Qa} to trade the gauge holonomies $a^I$ for the electric charges, and
\begin{equation}
(Qq) \,=\, Q_I q^I \,,\qquad (Q\tilde C^{-1}Q) \,=\, (\tilde C^{-1})^{IJ} Q_I Q_J\,.
\end{equation}

\paragraph{The entropy as a function of the charges.}  The above analysis gives three dipole charges $q^I$ related by the constraint~\eqref{eq:constr_dipole} fixing the dipole R-charge, three electric charges $Q_I$ related by the constraint~\eqref{eq:constraint_on_charges_0}, and two independent angular momenta $J_1$, $J_2$. 
It is not hard to find the combination of these quantities providing the Bekenstein-Hawking entropy \eqref{eq:BRBHE}. This requires using the relation \eqref{eq:gCtildeg} to eliminate $X$ from the first term in the expression~\eqref{eq:exprJ1J2_NH} for $J_1$, so that it can be rewritten as the square of the entropy~\eqref{eq:BRBHE}.
In this way one arrives at: 
\begin{equation} 
{\cal S} \,=\, \pi\, \sqrt{\frac{(Cqqq)\left(\big(Q\tilde C^{-1}Q\big)- \frac{\pi}{2G}J_1\right) + 3 \left( 2J_2+\left( Qq\right)\right)^2}{\big(g\tilde C^{-1} g\big)(Cqqq)} }\,.
\end{equation}
Remarkably, this matches precisely the expression \eqref{entropy_AdS_BR_bothsigma} for the entropy obtained from the Legendre transform of the on-shell action with $\sigma=-1$ upon using the map \eqref{eq:Ctok}.
Moreover, \eqref{eq:constraint_on_charges_0} agrees with the constraint on the electric charges obtained by imposing reality of the Legendre transform, given in~\eqref{eq:constraint_sigma_minus1}.

Finally, we note that using the expression for $J_2$ into the relation \eqref{eq:ratioomega_sigma-1} following from the Legendre transform of the action, we find that the horizon twist parameter is related to the angular potentials as $c = - \frac{\omega_2}{\omega_1}$. This implies that the supersymmetric Killing vector \eqref{eq:susyKV} can be expressed as the BTZ Killing vector
\begin{equation}
\xi \,=\, \frac{2\pi}{\beta}\partial_{\phi_0} + \frac{\ii\omega_1} {\beta} \partial_{\psi}\,,
\end{equation}
where we recall that $\omega_1$ is given by~\eqref{eq:omega1_maintext}.
The fact that there is no component along $S^2$ is consistent with the topological twist.

\section{Conclusions}\label{sec:conclusions}

In this work we proposed a systematic procedure for evaluating the on-shell action of five-dimensional supersymmetric black saddles with U(1)$^3$ isometry. This is made of three steps: (i) introduce a six-dimensional extension with the same fan as the five-dimensional geometry considered, (ii) identify a suitable resolution by decomposing the fan into simplicial cones to avoid singularities worse than orbifold, and (iii) localize the on-shell action, which takes the form of the anomaly polynomial six-form, over this space. The action integral reduces to a sum over contributions localized at the tip of each cone, corresponding to the fixed points of the action of the supersymmetric Killing vector. 

\paragraph{Checks and predictions.} For all cases where a comparison is possible, the results obtained by this method agree with the existing literature. In particular, in the ungauged case we reproduce the two-derivative on-shell action of explicitly known saddles with non-trivial horizon topology, recently computed in~\cite{Cassani:2025iix}, as well as the two-derivative expressions of~\cite{Colombo:2025yqy} for the action of putative asymptotically AdS$_5$ solutions that have not been explicitly constructed so far. Furthermore, we recover the known higher-derivative corrections to the action and to the extremal entropy of black holes with spherical horizon and either AdS or flat asymptotics~\cite{Cassani:2022lrk,Bobev:2022bjm,Cassani:2024tvk}, and also reproduce the known higher-derivative corrections to the entropy of asymptotically flat black rings~\cite{Guica:2005ig,Bena:2005ae}. The corrected entropies of the asymptotically flat black hole and black ring have recently been successfully compared with a microscopic index computation in~\cite{Alexandrov:2026rra}, which also indicates a black hole/black ring transition.

We have also proven a remarkable agreement between our expression for the entropy of the presumed asymptotically AdS$_5$ black ring with $\sigma=-1$, and the Bekenstein-Hawking entropy extracted from the explicit near-horizon geometry constructed in~\cite{Kunduri:2007qy}.
This agreement motivates further efforts towards constructing an explicit asymptotically AdS$_5$ black ring solution.
 
We also obtained new predictions for higher-derivative corrections, especially for black saddles with non-trivial horizon topology, both in asymptotically AdS$_5$ and asymptotically flat space. A compelling check would be to compare our formulae with a direct computation of higher-derivative corrections in the cases where the explicit (two-derivative) solution is known. For instance, for the asymptotically flat black lens such computation has not been carried out yet, but it should be achievable following the approach of~\cite{Cassani:2022lrk,Bobev:2022bjm}. 
At least in the asymptotically flat case, upon reducing along an isometry that preserves the Killing spinor, our results also relate with higher-derivative corrections to four-dimensional supersymmetric black hole saddles, see e.g.\ the recent discussions in~\cite{Hristov:2021qsw,Chen:2024gmc,Hegde:2024bmb,BenettiGenolini:2026qdm}.

\paragraph{Open questions.} Some issues remain to be addressed and more speculative aspects deserve further investigation. 
Firstly, it is necessary to verify if the higher-derivative on-shell action of supersymmetric solutions can indeed be recast in the form \eqref{eq:CS_HD}.
Extending the derivation of~\cite{Colombo:2025ihp,Colombo:2025yqy} to the four-derivative setting appears conceptually feasible, though  technically involved. Assuming this works as we have conjectured, it may be possible to extend the analysis to the full perturbative series in higher-derivative terms, showing that when evaluated on supersymmetric solutions with the prescribed asymptotics, terms with more than four derivatives can at most renormalize the coefficients $k_{IJK}$ and $k_I$, without affecting the rest of the on-shell action. See the discussions in \cite{Cassani:2025sim} and \cite{Hristov:2025ygn}. If this turns out to be correct, the black saddle on-shell actions given in the present paper would be exact in the full perturbative higher-derivative expansion.
 
For the asymptotically AdS black hole saddle, the on-shell action \eqref{eq:blackholeaction} is directly related to the Cardy-like asymptotics of the dual superconformal index. By contrast, for more general horizon topologies, the corresponding asymptotically AdS$_5$ solutions have not yet been constructed. It is therefore presently unclear whether the on-shell action formulae derived here are related to dual field theory saddles. If such AdS solutions do exist, and if they contribute to the gravitational index, we expect them to encode new Cardy-like asymptotics of the dual superconformal index.
 
Another intriguing question is whether the decomposition into simplicial cones of the toric fan associated with the solution admits a deeper mathematical or physical interpretation, beyond serving as a convenient tool for applying the localization formula. In particular, it would be highly interesting to determine whether this structure can provide new insights into the microstructure of composite black objects, such as black rings.

\paragraph{Relation to equivariant volume.} 
It was shown in \cite{Martelli:2023oqk,Colombo:2025yqy} that the on-shell action of a five-dimensional solution can be reorganized as an expansion whose terms are computed by the first four orders in the expansion of the equivariant volume of the associated six-dimensional toric geometry in powers of the K\"ahler parameters. While the agreement of the first three terms is automatic, the fourth and last term is more subtle, since additional manipulations are required in order to remove, or properly account for, the dependence on an auxiliary triangulation vector used to localize the equivariant volume. In our approach, instead, the integral of the anomaly polynomial over a closely related geometry gives the action directly, without requiring expansions or extra manipulations. It would be desirable to reach a deeper understanding of the relation between the two formalisms.

\paragraph{Extensions.} While we did not explicitly present examples with multiple horizons, such geometries can fall within the class described by the general formula~\eqref{eq:result_action} when preserving ${\rm U}(1)^3$ isometry, and would deserve a detailed study. Natural examples may be the concentric black rings of~\cite{Gauntlett:2004wh}, and their potential asymptotically AdS version.
A possible caveat is that, for more general solutions than the ones discussed here, the required resolution of the geometry may involve adding extra vectors to the original fan in order to obtain a good refinement.
One could also consider relaxing the assumption of U(1)$^3$ isometry and investigate whether analogous techniques can be applied to solutions with U(1)$^2$ symmetry.  Other possible generalizations include incorporating hypermultiplets and more general gaugings, as well as including solutions that are not necessarily asymptotically globally AdS$_5$ (or asymptotically flat in the ungauged limit), such as black strings. We expect our method to extend to these cases once a suitable reference background is identified.

\paragraph{Other dimensions and internal spaces.} A further extension would be to generalize our proposal to other dimensions. The asymptotically AdS$_3$ case seems very approachable. Recent work on asymptotically AdS$_7$ solutions that is close in spirit to the present paper can be found in~\cite{Hristov:2026tde}. One may even directly consider 10- or 11-dimensional supergravity, for instance type IIB supergravity compactified on five-dimensional Sasaki-Einstein manifolds. In the latter setting, fixed-point formulae can be applied both to the external space, along the lines of the present work, and to the internal manifold, since the volume of a Sasaki-Einstein space can be computed equivariantly by taking the cone over it. Hence, it is very natural to expect that the full on-shell (pseudo-)action of type IIB supergravity can be obtained by considering an integral over an auxiliary twelve-dimensional space and localizing to a simple sum over isolated fixed points. 

In the asymptotically flat case, there is a microscopic understanding of the gravitational index in terms of  wrapped M2-branes in Calabi-Yau geometries and Gopakumar-Vafa invariants \cite{Gopakumar:1998jq}.
It would be nice if the parallels between the asymptotically flat and asymptotically AdS$_5$ black saddles highlighted in the present paper extended beyond the similarities between saddle-point actions. This may suggest a relation between the corresponding gravitational indices, possibly along the lines of the giant graviton expansion~\cite{Imamura:2021ytr,Gaiotto:2021xce,Eleftheriou:2023jxr}.

We plan to come back to these questions in future work. 

\subsection*{Acknowledgments}

We acknowledge many useful discussions with Edoardo Colombo. We thank Alejandro Ruipérez for collaboration on related topics. ET is supported by the National Research Foundation of Korea under the grant RS-2025-00518906. We thank the Galileo Galilei Institute and the participants of the workshop ``Pathways to Quantum Black Holes'' for the stimulating environment that contributed to the development of part of this work. 

\appendix

\section{Wald entropy from Legendre transform}\label{appentropy}

\subsection{AdS$_5$ black ring}\label{app:WaldEntropyRings}

We consider four-derivative corrections to the candidate AdS$_5$ black ring. We start from the action~\eqref{eq:AdSBR_action}, which we reproduce here for convenience
\be
\label{eq:action_BR_appendix}
I(\omega, \varphi) \,=\, -k_{IJK}q^I\,\frac{3\varphi^J\varphi^K + 3\sigma\omega_2q^J\varphi^K+ \omega_2^2q^Jq^K}{6\sigma\omega_1} + k_I q^I\,\frac{\omega_1^2 + \omega_2^2+ \left(\omega_1 + \omega_2 - 2\varphi^{\cal R}\right)^2}{24\sigma\omega_1}\,,
\ee
where $\sigma^2 = 1$ and we used the constraint $\omega_1+\omega_2 - 2\varphi^{\cal R} = 2\pi \ii $.  The entropy is found by extremizing the following entropy function:
\begin{equation}
\label{eq:Legendre_BHBS}
{\cal S}={\rm ext}_{\{\varphi,\omega_1,\omega_2,\Lambda\}}\left[-I-\varphi^I Q_I-\omega_1 J_1-\omega_2 J_2 - \Lambda\left(\omega_1 + \omega_2 - 2\varphi^{\cal R} - 2\pi\ii\right)\right]\,.
\end{equation}
The extremization equations then read
\begin{equation}
\begin{aligned}
\frac{\partial I}{\partial \varphi^I} &= - \tilde k_{IJ} \frac{2\sigma\varphi^J + \omega_2 q^J}{2\omega_1} - \frac{\left(kq\right)}{6\sigma\omega_1}\left(\omega_1 + \omega_2 - 2\varphi^{\cal R}\right) g_I = - Q_I + 2\Lambda g_I\,,
\\[1mm]
\frac{\partial I}{\partial \omega_1}&= \tilde k_{IJ} \frac{3\varphi^I \varphi^J  +3\sigma \omega_2 q^I \varphi^J + \omega_2^2 q^I q^J}{6\sigma\omega_1^2} - \frac{\left(kq\right)}{24\sigma\omega_1^2}\left(\omega_2^2 -3\omega_1^2 - 2\omega_1\omega_2 + 4\omega_1\varphi^{\cal R}\right)
\\[1mm]
&= -\left(J_1 + \Lambda\right)\,,
\\[1mm]
\frac{\partial I}{\partial \omega_2} &= -\tilde k_{IJ} \frac{3q^I\varphi^J + 2\sigma\omega_2 q^I q^J}{6\omega_1} + \frac{\left(kq\right)}{12\sigma\omega_1}\left(\omega_1 + 2\omega_2 - 2\varphi^{\cal R}\right) = -\left(J_2 + \Lambda\right)\,,
\end{aligned}
\end{equation}
and satisfy the following equation 
\begin{equation}
\label{eq:BR_entropy0}
\begin{aligned}
&0 \simeq \left(kq\right) \frac{\left(kqqq\right)}{18} - \frac{2\sigma}{3}\left[ \left(kqqq\right) -\frac{3+\sigma}{2} \left(kq\right) \right] \frac{\partial I}{\partial \omega_1} + \left( q^I \frac{\partial I}{\partial \varphi^I}\right)^2 + 4 \left( \frac{\partial I}{\partial \omega_2} \right)^2+
\\[1mm]
&\frac{1}{3}\left[\left(kqqq\right) - \frac{3+\sigma}{2}\left(kq\right) \right] \big(\tilde k^{-1}\big)^{IJ}\frac{\partial I}{\partial \varphi^I}\frac{\partial I}{\partial \varphi^J}  - 4\sigma q^I \frac{\partial I}{\partial \varphi^I}\frac{\partial I}{\partial \omega_2} +\left(kq\right) \frac{1+\sigma}{6}\left[q^I \frac{\partial I}{\partial \varphi^I} - 2\sigma\frac{\partial I}{\partial \omega_2}\right] 
\\[1mm]
&+ \left(kq\right) \sigma g_I \big(\tilde k^{-1}\big)^{IJ} \frac{\partial I}{\partial \varphi^J}\left[\frac{\left(kqqq\right)}{9}\left( 1 + \sigma g_I \big(\tilde k^{-1}\big)^{IJ} \frac{\partial I}{\partial \varphi^J}\right) +\frac{1+\sigma}{3} \left(\sigma q^I \frac{\partial I}{\partial \varphi^I} - 2\frac{\partial I}{\partial \omega_2}\right)\right]\,.
\end{aligned}
\end{equation}
To write these expressions we have imposed $q_1^{\cal R} =(1-\sigma)/2$, which follows from \eqref{eq:BR_constraint}, and employed the same notation introduced around \eqref{eq:notation_LT}. Moreover, the symbol ``$\simeq$'' denotes that we are neglecting all terms of order ${\cal O}\left( k_I^2\right)$.
Since the action \eqref{eq:action_BR_appendix} is homogeneous of degree $1$ in the chemical potentials, as a consequence of Euler's theorem for homogeneous functions we find,
\begin{equation}
{\cal S} = 2\pi \ii \Lambda\,,
\end{equation}
where the extremal value of $\Lambda$ solves the quadratic equation that follows from \eqref{eq:BR_entropy0}:
\begin{equation}\label{eq:quadraticeq}
P_0 + P_1 \Lambda + P_2 \Lambda^2 \simeq 0\,,
\end{equation}
with
\begin{equation}
\begin{aligned}
P_2 &= \frac{4}{3}\left(g\tilde k^{-1} g\right)\left(kqqq\right) + 2\left(\sigma +1 \right)+ \frac{2\left(kq\right)}{9} \left(g\tilde k^{-1}g\right) \left[2\left(kqqq\right)\left(g\tilde k^{-1}g\right) + 3\left(\sigma-1\right)\right] \,,
\\[2mm]
P_1 &=  \frac{2}{3} \left[ \left( \sigma -2 \left(g\tilde k^{-1}Q\right) \right) \left( kqqq\right) + 3 \left(1+\sigma\right)\Bigl(2 J_2 - \left(Qq\right)\Bigr)  \right]- \frac{2\left( kq\right)}{9}  \Biggl[3\sigma
\\[1mm]
&\quad - \left(g\tilde k^{-1}g\right) \Bigl( \sigma \left(kqqq\right) + 3\left(1+\sigma\right) \Bigl(2J_2-\left(Qq\right)\Bigr)\Bigr) -2\left(g\tilde k^{-1}Q\right)\left(3-\left(g\tilde k^{-1}g\right)\left(kqqq\right)\right)\Biggr]\,,
\\[2mm]
P_0 &= \left(Qq\right)^2 + 4J_2\Bigl( J_2 - \sigma\left(Qq\right)\Bigr) + \frac{\left(kqqq\right)}{3}\Bigl(Q\tilde k^{-1} Q + 2\sigma J_1\Bigr) + \frac{\left(kq\right)}{18}\Biggl[
\left(kqqq\right) \Bigl( 1 - 2\sigma \left(g\tilde k^{-1} Q\right) 
\\[1mm]
&\quad + 2 \left( g\tilde k^{-1} Q\right)^2 \Bigr) - 3\left( 3+ \sigma\right) \left( Q\tilde k^{-1} Q\right) +3\left( 1 + \sigma\right)  \Bigl(1-2\left(g\tilde k^{-1} Q\right) \Bigr) \Bigl( 2 J_2 - \left( Qq\right) \Bigr)
\\[1mm] 
&\quad - 6\left( 1 + 3\sigma\right) J_1 
\Biggr]\,,
\end{aligned}
\end{equation}
where $
\big(B\tilde k^{-1}C\big) \equiv \big(\tilde k^{-1}\big)^{IJ} B_I C_J\,.$

The entropy is therefore given by
\begin{equation}
\label{eq:entropy_BHBS_complex}
{\cal S} \,\simeq\, \pi \sqrt{4P_2^{-1}P_0 - (P_2^{-1}P_1)^2} - \ii \pi P_2^{-1}P_1\,.
\end{equation}
For real charges, the physical entropy of any Lorentzian solution connected with the Euclidean geometry should be real. Requiring the imaginary part of
\eqref{eq:entropy_BHBS_complex} to vanish then imposes the charge constraint:\footnote{
A second possibility to obtain a real entropy would be to impose that $P_0 \simeq 0$~\cite{Cassani:2024kjn,Cassani:2025iix}. After imposing this constraint, the entropy would be vanishing, ${\cal S}=0$. The corresponding geometry satisfying such constraint, then, is expected to be an asymptotically AdS$_5$ horizonless topological soliton, which carries zero entropy, generalizing the one studied in~\cite{Chong:2005hr}.
}
\begin{equation}
P_1 \simeq 0\,.
\end{equation}
Then, the entropy of the AdS$_5$ black rings -- at linear order in the four-derivative corrections controlled by the linear anomaly coefficient -- can be expressed as
\begin{equation}
{\cal S} \simeq 2\pi\sqrt{P_2^{-1}\,P_0}\,.
\end{equation}
In particular, we observe that the entropy \eqref{eq:AFBR_entropy} and charge constraint \eqref{eq:AFBR_charges} for the asymptotically flat black ring are reproduced from the expressions in this section by taking $\sigma =1$ together with the ungauged limit $g_I =0$.

The Bekenstein-Hawking entropy and the corresponding leading-order charge constraint can be found by setting to zero the corrections controlled by $k_I$. 

For $\sigma =1$, we find that the charge constraint can be solved for $J_2$, 
\begin{equation}
P_1 = 0\quad \implies\quad J_2 = \frac{\left(Qq\right)}{2} - \frac{\left(kqqq\right)}{12}\left(1-2\big(g\tilde k^{-1}Q\big)\right)\,,
\end{equation}
and the entropy reduces to
\begin{equation}
{\cal S} \,=\, \pi \sqrt{\frac{ \left(kqqq\right) \left( \big(Q\tilde k^{-1} Q\big) +2 J_1 \right) + 3\left( \left( Qq\right) - 2 J_2\right)^2 }{\big(g\tilde k^{-1}g \big) \left(kqqq\right) + 3}}\,,
\end{equation}
reproducing the results of~\cite{Colombo:2025yqy}.

For $\sigma =-1$, besides~\eqref{eq:quadraticeq} the extremization equations imply
\be
\varphi^I \,=\, - \omega_1  (\tilde k^{-1})^{IJ}(Q_J - 2\Lambda g_J) + \frac{1 }{2}\omega_2\, q^I  \,,
\ee
\be\label{eq:ratioomega_sigma-1}
\frac{\omega_2}{\omega_1} \,=\, - 6\,\frac{(Qq)+2J_2}{(kqqq)}\,.
\ee
Contracting the first equation with $2g_I$, using the constraint between the chemical potentials as well as ${\rm Im}\,\mathcal{S} = \frac{\pi}{2}\frac{1+2(g\tilde k^{-1}Q)}{(g\tilde k^{-1}g)}$, we can solve for $\omega_1$,
\be\label{eq:expr_omega1_sigma-1}
\omega_1\,=\, \frac{\pi^2}{{\rm Re}\,\mathcal{S} \,(g\tilde k^{-1} g)}\,.
\ee
The charge constraint gives the following relation
\begin{equation}
P_1 = 0\quad \implies\quad \big( g\tilde k^{-1} Q\big) = - \frac{1}{2}\,.
\end{equation}
We observe that the same value of $\big( g\tilde k^{-1} Q\big) $ solves the charge constraint also after turning on higher-derivative corrections (at least at linear order). 
The Bekenstein-Hawking entropy for this branch of the solution is  given by
\begin{equation}
{\cal S} \,=\, \pi \sqrt{\frac{ \left(kqqq\right) \left( \big(Q\tilde k^{-1} Q\big) -2 J_1 \right) + 3\left( \left( Qq\right) + 2 J_2\right)^2 }{\big(g\tilde k^{-1}g \big) \left(kqqq\right)}}\,.
\end{equation}

\subsection{Black hole in bubbling spacetime}
\label{app:BHinBS}

In this appendix, we derive the Wald entropy of a black hole in the background of a topological soliton, in the ungauged case. Our result is consistent with the Bekenstein-Hawking entropy obtained in \cite{Horowitz:2017fyg} by studying the corresponding (explicit and Lorentzian) extremal asymptotically flat solution.

We focus on the limit to minimal ungauged supergravity. This gives the following action: 
\begin{equation}
\label{eq:action_BHBS_minimal}
\begin{aligned}
I&=\frac{\kappa}{6}\,\left[\frac{\varphi^3}{\omega_1\omega_2} + q_2\frac{3\left(\varphi + \omega_1 q_3\right)^2 + 3 \omega_2q_2\left(\varphi + \omega_1q_3\right) + \omega_2^2 q_2^2}{\omega_1} \right] 
\\[1mm]
&\quad - \frac{k_1}{24}\left(\varphi + \omega_2q_2\right) \frac{\omega_1^2 + \omega_2^2 +\left(\omega_1 + \omega_2\right)^2}{\omega_1\omega_2}\,,
\end{aligned}
\end{equation}
where we denote $\kappa \equiv k_{111}$ and we have used the constraint $\omega_1 + \omega_2 = 2\pi\ii$.
We need to extremize the entropy function:
\begin{equation}
\label{eq:Legendre_BHBS}
{\cal S}={\rm ext}_{\{\varphi,\omega_1,\omega_2,\Lambda\}}\left[-I-\varphi Q-\omega_1 J_1-\omega_2 J_2 - \Lambda\left(\omega_1 + \omega_2 - 2\pi\ii\right)\right] = 2\pi \ii \Lambda\,.
\end{equation}
The extremal value of $\Lambda$ solves the quadratic equation
\begin{equation}
\label{eq:quadr_lambda}
P_0 + P_1 \Lambda + \Lambda^2 \simeq 0\,,
\end{equation}
with coefficients
\begin{equation}
\begin{aligned}
P_1 &= J_1 + J_2 - q_2 Q - \frac{\kappa}{2}q_2 q_3\left(q_2-q_3\right) + \frac{k_1}{2}\frac{J_1-J_2}{Q + \kappa\, q_2 q_3}\,,
\\[1mm]
4P_0 &= \left(1-\frac{3k_1}{4\left(Q+\kappa\, q_2 q_3\right)}\right)\Biggl(
\frac{8}{9\kappa}\left[Q+\kappa\,q_2q_3\right]^3
\\[1mm]
&\quad -\left( 1 - \frac{k_1}{4\left(Q + \kappa\,q_2q_3\right)}\right)\Bigl[J_1-J_2 + q_2 Q+\frac{\kappa}{2}q_2q_3\left(q_2+q_3\right)\Bigr]^2
\Biggr)
\\[1mm]
&\quad+\frac{k_1}{2}\left(J_1 - J_2\right) \frac{q_2\left(2Q + \kappa \,q_3\left(q_2-q_3\right)\right)}{Q + \kappa \,q_2 q_3}+P_1^2\,.
\end{aligned}
\end{equation}
As above, the symbol ``$\simeq$'' in \eqref{eq:quadr_lambda} denotes the fact that we are working at linear order in the corrections, i.e. all terms of order ${\cal O}\left( k_1^2\right)$ are neglected. To obtain a real entropy the following constraint needs to be imposed:\footnote{As above, a second possibility would be to impose $P_0 \simeq 0$. This second case provides a vanishing entropy, corresponding to the constraint associated with a horizonless soliton, including higher-derivative corrections at linear order.}
\begin{equation}
P_1 = J_1 + J_2 - q_2 Q - \frac{\kappa}{2}q_2 q_3\left(q_2-q_3\right) + \frac{k_1}{2}\frac{J_1-J_2}{Q + \kappa \,q_2 q_3}\simeq 0\,.
\end{equation}
After imposing this, the entropy at linear order in the corrections reduces to
\begin{equation}
\begin{aligned}
{\cal S} \,&\,\simeq  \pi \Biggl[\left(1-\frac{3k_1}{4\left(Q+\kappa \,q_2 q_3\right)}\right)\Biggl(
\frac{8}{9\kappa}\left[Q+\kappa\,q_2q_3\right]^3
\\[1mm]
&\,\quad -\left( 1 - \frac{k_1}{4\left(Q + \kappa\,q_2q_3\right)}\right)\Bigl[J_1-J_2 + q_2 Q+\frac{\kappa}{2}q_2q_3\left(q_2+q_3\right)\Bigr]^2
\Biggr)
\\[1mm]
&\,\quad+\frac{k_1}{2}\left(J_1 - J_2\right) \frac{q_2\left(2Q + \kappa \,q_3\left(q_2-q_3\right)\right)}{Q + \kappa \,q_2 q_3}\Biggr]^{1/2}\,.
\end{aligned}
\end{equation}
The Bekenstein-Hawking entropy is the leading-order expression obtained by taking $k_1 =0$:
\begin{equation}
\label{eq:entropy_BHBS}
\begin{aligned}
{\cal S}&=2\sqrt{\pi}\sqrt{\frac{1}{6\sqrt{3}}\left(Q+\frac{16\pi}{\sqrt{3}}q_2q_3\right)^3-\pi\left(J_- + J_+ + \frac{16\pi}{\sqrt{3}}q_2q_3^2\right)^2}\,,
\end{aligned}
\end{equation}
with charges constrained by
\begin{equation}
\label{eq:constraint_BHBS}
J_+-q_2 Q-\frac{8\pi}{\sqrt{3}}q_2q_3(q_2-q_3)=0\,,
\end{equation}
where $J_\pm = \frac{J_1 \pm J_2}{2}$. To write \eqref{eq:entropy_BHBS} and \eqref{eq:constraint_BHBS} we have employed the following dictionary
\begin{equation}
\kappa = \frac{3\pi}{2}\,,\qquad Q\to \frac{\sqrt{3}}{2}Q\,,\qquad q_{2,3} \to \frac{4}{\sqrt{3}}q_{2,3}\,.
\end{equation}
With these choices it is now manifest that the charge constraint \eqref{eq:constraint_BHBS} and the entropy \eqref{eq:entropy_BHBS} reproduce the results of \cite{Horowitz:2017fyg}. This provides a non-trivial consistency check of our findings.

\bibliography{equiv_int.bib}
\bibliographystyle{JHEP}

\end{document}